\documentclass[reprint,superscriptaddress,amsmath,amssymb,
 aps,showkeys]{revtex4-2}
\usepackage{graphicx}
\usepackage{dcolumn}
\usepackage{bm}
\usepackage{multirow}
\usepackage{booktabs}
\usepackage{float}
\usepackage{placeins}
\usepackage{xcolor}
\usepackage[colorlinks=true,linkcolor=blue,citecolor=blue,urlcolor=blue]{hyperref}
\usepackage{miller}

\begin{document}

\preprint{APS/123-QED}

\title{Grain boundary segregation spectrum in basal-textured Mg alloys: \\From solute decoration to structural transition} 

\author{Anumoy Ganguly}
\affiliation{Institute of Physical Metallurgy and Materials Physics, RWTH Aachen University, 52056 Aachen, Germany}

\author{Hexin Wang}
\affiliation{Institute of Physical Metallurgy and Materials Physics, RWTH Aachen University, 52056 Aachen, Germany}

\author{Julien Guénolé}
\affiliation{Université de Lorraine, CNRS, Arts et Métiers, LEM3, 57070 Metz, France}

\author{Aruna Prakash}
\affiliation{Micro‑Mechanics and Multiscale Materials Modeling (M5), Institute of Mechanics and Fluid Dynamics, TU Bergakademie Freiberg, 09599 Freiberg, Germany}
     
\author{Sandra Korte-Kerzel}
\affiliation{Institute of Physical Metallurgy and Materials Physics, RWTH Aachen University, 52056 Aachen, Germany}

\author{Talal Al-Samman}
\affiliation{Institute of Physical Metallurgy and Materials Physics, RWTH Aachen University, 52056 Aachen, Germany}

\author{Zhuocheng Xie}
\email[]{xie@imm.rwth-aachen.de}
\affiliation{Institute of Physical Metallurgy and Materials Physics, RWTH Aachen University, 52056 Aachen, Germany}

\begin{abstract}
Mg alloys are promising lightweight structural materials due to their low density and excellent mechanical properties. However, their limited formability and ductility necessitate improvements in these properties, specifically through texture modification via grain boundary segregation. While significant efforts have been made, the segregation behavior in Mg polycrystals, particularly with basal texture, remains largely unexplored. In this study, we performed atomistic simulations to investigate grain boundary segregation in dilute and concentrated solid solution Mg-Al alloys. We computed the segregation energy spectrum of basal-textured Mg polycrystals, highlighting the contribution from specific grain boundary sites, such as junctions, and identified a newly discovered bimodal distribution which is distinct compared to the conventional skew-normal distribution found in randomly-oriented polycrystals. Using a hybrid molecular dynamics/Monte Carlo approach, we simulated segregation behavior at finite temperatures, identifying grain boundary structural transitions, particularly the varied fraction and morphology of topologically close-packed grain boundary phases when changing thermodynamic variables. The outcomes of this study offer crucial insights into basal-textured grain boundary segregation and phase formation, which can be extended to other relevant Mg alloys containing topologically close-packed intermetallics.
\end{abstract}


\maketitle

\renewcommand{\figurename}{Figure}
\renewcommand{\tablename}{Table}

\section{\label{Intro}Introduction}

Magnesium alloys, known for their excellent strength-to-weight ratio, are emerging as promising lightweight materials for enhancing sustainable low-carbon automotive applications \cite{mordike2001magnesium,pollock2010weight}. Nevertheless, the broader commercial use of Mg alloy sheets faces substantial challenges due to considerable processing difficulties, primarily caused by their anisotropic plasticity and poor formability at ambient temperature. These limitations arise from the restricted number of available deformation modes at low temperatures and the strong basal texture, which hinders through-thickness deformation of Mg sheets during forming \cite{yoo1981slip,kocks2000texture}. In recent years, considerable efforts have been made to improve the formability of Mg alloys and further enhance their strength. These efforts include targeted addition of alloying elements to minimize the differences in critical resolved shear stresses between basal and non-basal slip systems \cite{sandlobes2011role,ovri2023mechanistic}, and to weaken the strong basal texture through grain boundary (GB) segregation \cite{basu2014triggering,trang2018designing}.

Solute segregation at GBs in Mg alloys with dilute solute additions has been extensively documented as a factor influencing GB energy and mobility, which in turn affects the characteristics of GB migration during annealing, and the resulting texture development \cite{zeng2016texture,barrett2017effect}. Understanding the atomistic mechanisms behind GB segregation is essential for designing Mg alloys with customized properties.
Solute decoration at GBs in Mg alloys was reported by Nie et al. \cite{nie2013periodic}, who identified periodic substitutional segregation of Gd and Zn solutes at twin boundaries using high-resolution scanning transmission electron microscopy (HR-STEM). They attributed this periodic solute decoration to strain energy minimization in the matrix, as confirmed by density functional theory (DFT) calculations. The resulting pinning effect from ordered solute segregation at twin boundaries can significantly enhance the strengthening of Mg alloys. Similar atomic-resolution electron microscopy approaches have been applied to study other highly symmetric tilt GBs in Mg alloys, revealing characteristic solute decoration patterns, which further underscores the role of solute segregation in influencing the structural properties of grain boundaries \cite{zhou2015effect,he2021unusual,xie2021nonsymmetrical}.

Atomic-scale modeling techniques, including DFT and atomistic simulations, have enhanced our understanding of GB segregation across a wide range of composition space. Huber et al. \cite{huber2014atomistic} calculated the per-site segregation energy $\Delta E_{\text{seg}}$ of 11 alloying elements at $\Sigma$7 GBs in Mg using DFT, and developed a linear elastic model to predict $\Delta E_{\text{seg}}$ based on the GB site volume. Wang et al. \cite{wang2024defects} assessed Mg binary semi-empirical potentials for $\Delta E_{\text{seg}}$ of various alloying elements at highly symmetric GBs in Mg, finding qualitative agreement with DFT calculations \cite{huber2014atomistic,pei2019first}. Messina et al. \cite{messina2021machine} employed machine learning algorithms to correlate $\Delta E_{\text{seg}}$ of Al solute at Mg \hkl<0001> symmetric tilt GBs calculated using atomistic simulations to the structural and energetic descriptors. Pei et al. \cite{pei2023atomistic} investigated the segregation behavior of Nd solute at general Mg GBs using atomistic simulations and atom probe tomography (APT). The study found that the inhomogeneous segregation behavior within the GB plane stems from local atomic arrangements within the GBs rather than macroscopic characteristics. Recently, Menon et al. \cite{menon2024atomistic} calculated $\Delta E_{\text{seg}}$ of Y solutes at symmetric tilt GBs in Mg using atomistic simulations and estimated the per-site segregation free energy at finite temperatures. The predicted GB concentrations of Y solutes showed a good correlation with experimental measurements at typical processing temperatures in Mg alloys, based on the classical Langmuir–McLean isotherm \cite{mclean1957grain} and its extension by White and Coghlan \cite{white1977spectrum} to account for the spectrality of segregation at the GB site level.

For the investigation of GB segregation beyond the dilute limit, solute-solute interactions need to be considered when predicting the GB concentrations, i.e., using the Fowler-Guggenheim isotherm, which incorporates a solute interaction term \cite{fowler1939statistical,lejcek2010grain}.
In this context, strong attractive interactions lead to solute clustering and preferential co-segregation in the vicinity of defects, such as dislocations and GBs, and affect their mobilities \cite{stanford2009atom,bugnet2014segregation,robson2016grain,bian2018bake,pei2022effect,mouhib2022synergistic}.
Recently, Mouhib et al. \cite{mouhib2024exploring} used APT to characterize the GB concentrations of Ca and Gd solutes in rolled and annealed Mg alloys, both with and without the addition of Zn. They found that the selective formation of specific texture components in the ternary Mg-Zn-X alloys (X=Ca or Gd) was influenced by the synergistic effects of Zn addition, promoting solute binding and co-segregation.

From the perspective of enhancing Mg alloy strength through aging at low temperatures, it is important to understand the mechanisms involving solid solution decomposition and the formation of secondary phases, capable of effectively hindering dislocation slip and twinning growth \cite{nie2012precipitation}. In regard to the age-hardening response of Mg-Al alloys, Clark \cite{clark1968age} reported that the precipitation of the equilibrium $\beta$-Mg\textsubscript{17}Al\textsubscript{12} phase can be both discontinuous (at the GBs) and continuous (in the grain interior), giving rise to cellular and plate-like precipitates, respectively. Further work demonstrated that controlling the aging temperature between 423 K and 693 K determines whether discontinuous, continuous or both forms of precipitation occur simultaneously \cite{braszczynska2009discontinuous}.
Yang et al. \cite{yang2018precipitation} reported that Zn atoms segregate at cores of prismatic dislocations in cold-rolled and annealed Mg-5Zn and Mg-9Zn (wt.\%) alloys, forming incomplete icosahedral chains, similar to the C14 MgZn\textsubscript{2} structure, along the dislocation lines, as observed using HR-STEM. Under higher annealing temperatures, larger quasicrystalline precipitates formed, exhibiting more intricate arrangements, including C14 and C15 Laves phases, as well as other topologically close-packed (TCP) structures with multiple orientations. The segregation of Zn along dislocations and the subsequent atomic arrangement leading to the structural transformation into quasicrystalline patterns were identified as a key mechanism for precipitation strengthening using quasicrystals. 

To date, most atomistic studies on GB segregation in Mg alloys have primarily focused on individual GBs, particularly highly symmetric tilt GBs. While the GB segregation energy spectrum and associated local atomic environment space have been extensively investigated by Wagih and Schuh et al. \cite{wagih2019spectrum,wagih2020learning,wagih2022learning,wagih2023can} for randomly-textured FCC polycrystals, similar research on textured polycrystals is lacking. This is particularly true for polycrystalline Mg with strong basal texture, commonly found in conventional sheet alloys. In addition, the mechanisms of GB structural transitions due to solute segregation at finite temperatures, especially the nucleation and growth of GB phases in Mg alloys, remain unknown. In this study, we performed atomistic simulations to compute the segregation energy spectrum of basal-textured polycrystals in Mg-Al alloys and compared it with that of randomly-oriented polycrystals and symmetric tilted GBs. We simulated segregation behavior at finite temperatures using a hybrid molecular dynamics/Monte Carlo (MD/MC) approach and investigated the GB structural transitions under different thermodynamic variables.

\section{\label{Methods}Simulation methods}

\subsection{Atomistic simulations}

Atomistic simulations in this study were performed using the MD software package LAMMPS~\cite{thompson2022lammps}. 
Interatomic interactions were modeled by the Finnis-Sinclair embedded atom method (EAM) potentials developed by Mendelev et al. for Mg-Al \cite{mendelev2009development}. This potential accurately predicts Mg bulk properties, including elastic constants and stacking fault energies, as well as per-site segregation energies of Al solutes at Mg GBs, showing good agreement with experimental and ab-initio results \cite{wang2024defects}. 

In total, 18 \hkl<0001> symmetric tilt coincidence site lattice (CSL) GB configurations with $\Sigma<100$ ($\Sigma$7, $\Sigma$13, $\Sigma$31, $\Sigma$37, $\Sigma$61, $\Sigma$67, $\Sigma$73, $\Sigma$79, $\Sigma$97) were constructed using Atomsk~\cite{hirel2015atomsk} as illustrated in  \cite{zhang2022atomistic}. A Mg unit cell oriented along $x$-\hkl[2 -1 -1 0], $y$-\hkl[0 1 -1 0], and $z$-\hkl[0 0 0 1] was rotated around the $z$-axis by half of the tilt angles. The periodicity of the supercell was maintained in all directions post-rotation. Two symmetrically rotated crystals were superimposed along $x$, and semi-fixed boundary conditions were applied at both ends of the sample, each with a thickness of two times the potential cutoff (2 $\times$ 6~\AA). The distance from the GB plane to the ends is approximately 20~nm to minimize the interactions between the GB and the semi-fixed boundaries. Periodic boundary conditions were applied along the $y$ and $z$ directions. To explore the microscopic degrees of freedom of the GB structures, rigid body translations in all directions and atom deletion within different cutoff distances ranging from 0.6 to 2~\AA were implemented. Relaxation perpendicular to the GB plane was conducted after each translation step using the FIRE algorithm~\cite{bitzek2006structural,guenole2020assessment}. Full relaxation of the minimum energy configurations was achieved using both the conjugate gradient (CG) and FIRE algorithms with a force tolerance of $10^{-8}$~eV/\AA.

\begin{figure*}[hbt!]
\centering
\includegraphics[width=0.75\textwidth]{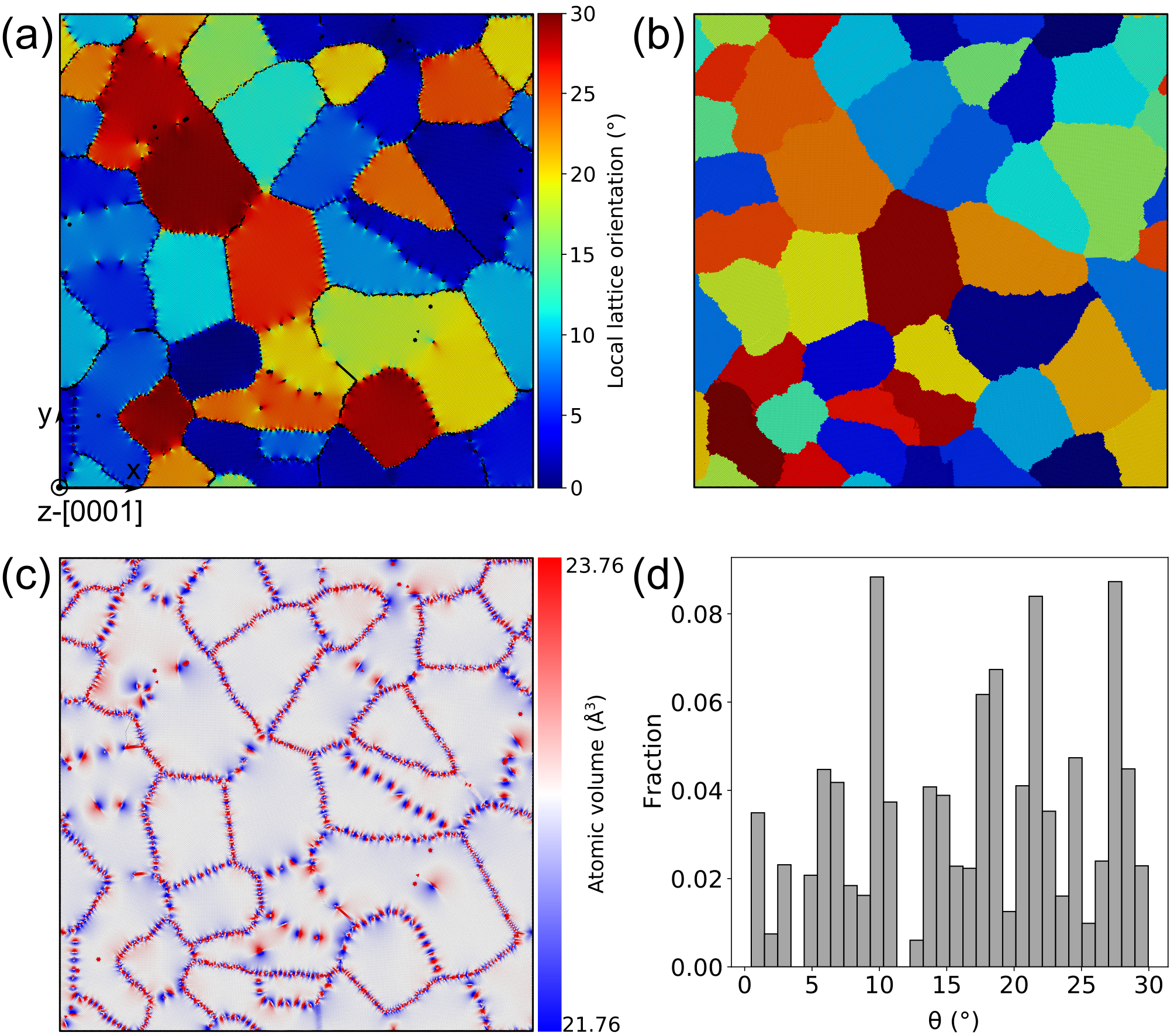}
\caption{Atomistic configuration of the studied basal-textured polycrystal with an average grain size ($D$) of 17.8~nm. (a) Orientation variation of the basal plane within the $x$-$y$ plane via the PTM algorithm, with GB sites (non-HCP type) colored black. Here 0° corresponds to $x$-\hkl[2 -1 -1 0], and 30° to $y$-\hkl[-1 -1 2 0]. (b) Grain segmentation via the Orisodata algorithm. (c) Atomic volume via the Voronoi tessellation. The median value of the color bar represents the atomic volume in bulk Mg ($V_{bulk}^{i}=22.76$~\AA$^{3}$). (d) GB  misorientation angle ($\theta$) distribution of the segmented microstructure for GB sites with more than two neighboring grains within a cutoff radius of 7.8~\AA. The orientation of each grain was defined by the average local lattice orientation in the grain via PTM.} 
\label{fig1}
\end{figure*}

Polycrystalline Mg structures featuring columnar grains oriented in the $z$-\hkl[0001] direction and with a thickness of 3.13~nm were constructed using Atomsk, based on Voronoi tessellation (Figure~\ref{fig1}). The periodic simulation box, measuring 100~nm in the $x$ and $y$ dimensions, contains approximately 1,360,000 atoms. The basal-textured polycrystals consist of 20, 40, 100, and 400 grains, with average grain sizes ($D$) of 25.2, 17.8, 11.3, and 5.6 nm calculated using average spherical equivalent grain diameter. Using the polyhedral template matching (PTM) approach~\cite{larsen2016robust}, 3.1\%, 4.4\%, 6.5\%, and 12.8\% of the sites in the generated structures are identified as GB sites (Figure~\ref{fig1}(a)). For comparison, randomly-oriented polycrystals with dimensions of 10 $\times$ 10 $\times$ 10~nm$^{3}$ containing 5 and 10 grains with average grain sizes of 7.2 and 5.8~nm, respectively, were created.
Following initial energy minimization, the polycrystalline sample was heated to 600~K over 60~ps and kept at that temperature for 50~ps before being quenched back to 0~K over 60~ps using the isothermal-isobaric ensemble with the Nosé-Hoover thermostat and barostat~\cite{hoover1985canonical}. The timestep is 1~fs. Subsequently, the FIRE algorithm was applied to relax the structure with a force tolerance of $10^{-8}$~eV/\AA.

\subsection{Per-site segregation energy and solute concentration}

The per-site segregation energy $\Delta E_{\text{seg}}^{i}$ at 0~K for a substitutional solute Al atom at a GB site $i$ characterized using PTM was calculated using Equation~\ref{equ1}:
\begin{equation}
\label{equ1}
\Delta E_{\text{seg}}^{i} = (E_{\text{bulk}} - E_{\text{bulk,X}}) - (E_{\text{GB}} - E_{\text{GB,X}}^{i}),
\end{equation}
where $E_{\text{bulk}}$ represents the energy of a bulk system, $E_{\text{bulk,X}}$ denotes the energy of the bulk system with a host atom replaced by an X solute, $E_{\text{GB}}$ is the energy of a system containing a GB. After the solute substitution, the FIRE algorithm was applied to relax the solute-decorated GB structure with a force tolerance of $10^{-8}$~eV/\AA. $E_{\text{GB,X}}^{i}$ stands for the energy of the relaxed system with a solute atom X occupying the GB site $i$. A negative segregation energy indicates favorable solute segregation.

The per-site segregation energy, $\Delta E_{\text{seg}}^{i}$, can be directly correlated with solute concentration at GBs based on the Langmuir–McLean isotherm and the White-Coghlan extension, as illustrated in Equation~\ref{equ2} in the context of an infinitesimally dilute solid solution: 
\begin{equation}
\label{equ2}
X_{\text{GB}}^{i} = \left(1 + \frac{1 - X_{\text{bulk}}}{X_{\text{bulk}}} \exp\left(\frac{\Delta E_{\text{seg}}^{i}}{k_B T}\right)\right)^{-1},
\end{equation}
where $X_{\text{bulk}}$ is the solute concentration in the bulk region, and $X_{\text{GB}}^{i}$ is the fraction of solute concentration of a GB site $i$. This approach takes into account the spectrum of solute concentrations found in the polycrystal, allowing calculation of the weighted average GB solute concentration ($X_\text{GB}$) from Equation~\ref{equ3}: 
\begin{equation}
\label{equ3}
X_\text{GB} = \sum_i F^{i} X_{\text{GB}}^{i},
\end{equation}
where $F^{i}$ is the fraction of GB sites with the per-site segregation energy $\Delta E_{\text{seg}}^{i}$. Such a methodological approach allows for direct comparison and correlation of theoretical predictions from atomic-scale modeling with experimentally measured solute concentrations at GBs.

\subsection{Hybrid MD/MC simulations}

The polycrystalline Mg configurations were equilibrated for 200~ps at 300~K or 600~K using the Nosé-Hoover thermostat and barostat to relax possible stresses. To investigate the segregation behavior of Al solutes at finite temperatures, hybrid MD/MC simulations in the variance-constrained semi-grand canonical (VC-SGC) \cite{sadigh2012scalable} ensemble were performed at 300 and 600~K for at least 2 and 1~ns, respectively (1 MC simulation step with a 10\% swap fraction every 0.1~ps MD simulation time with an MD time step of 1~fs). The trial moves are accepted with probability $A$ defined in Equation~\ref{equ4}:
\begin{equation}
\label{equ4}
\begin{aligned}
A = & \min\left(1, \exp\left\{-\frac{\Delta U + N\Delta c(\phi + 2\kappa N c_{0})}{k_B T}\right\}\right) \\
  & \times \min\left(1, \exp\left\{-\frac{ \kappa N^{2} \Delta c[\Delta c - 2 (\hat{c} - c_{0})]}{k_B T}\right\}\right),
\end{aligned}
\end{equation}
where $\Delta U$ and $\Delta c$ are the changes in the potential energy and concentration due to the trial move. The concentration before the trial move is $\hat{c}$. The chemical potential difference $\Delta\mu_{0}=\phi + 2\kappa N c_{0}=1.88$~eV is defined as the difference in chemical potentials of HCP Mg and FCC Al. The $\kappa$ value is 1000. The target concentrations $c_{0}$ of Al atoms range from 0.1 to 10\%. Equilibrium at the target concentrations and temperatures was monitored by observing the evolution of potential energies and potential energy differences (see Figure S1-6). To minimize thermal noise for subsequent structural analysis and visualization, a quenching process using the CG algorithm for 50 steps was performed following the hybrid MD/MC simulations.

\subsection{Structural analysis and visualization}

The grain segmentation (Figure~\ref{fig1}(b)) within the basal-textured polycrystal (Figure~\ref{fig1}(a)) was performed using the Or\textsc{isodata} algorithm \cite{vimal2022grain} (for the parameters see Table SI). GB sites were classified into two categories: those located at the GB plane shared by two neighboring grains and those at junctions shared by more than two neighboring grains. Additionally, GB plane sites were further categorized into low-angle and high-angle GB sites, based on the misorientation angles of neighboring grains around the \hkl[0001] axis (Figure~\ref{fig1}(d)). These angles were determined by taking the average local lattice orientation of the adjacent grains obtained via PTM.
After the hybrid MD/MC simulations, the polycrystalline structures with GB solute segregation were analyzed using PTM to assess structural and chemical ordering. Voronoi tessellation was employed to calculate the atomic volume and identify Frank-Kasper atomic clusters, indicative of TCP phases. The presence of Laves phases was confirmed by the LaCA algorithm \cite{xie2021laves}. Local entropy \cite{piaggi2017entropy} was used to characterize the ordering of the polycrystalline structures. The Open Visualization Tool (OVITO) \cite{Stukowski_2010} was used to visualize the atomic configurations. 

\section{\label{Results}Results}

\subsection{\label{Results1}Segregation in dilute solid solution}

\begin{figure*}[hbt!]
\centering
\includegraphics[width=\textwidth]{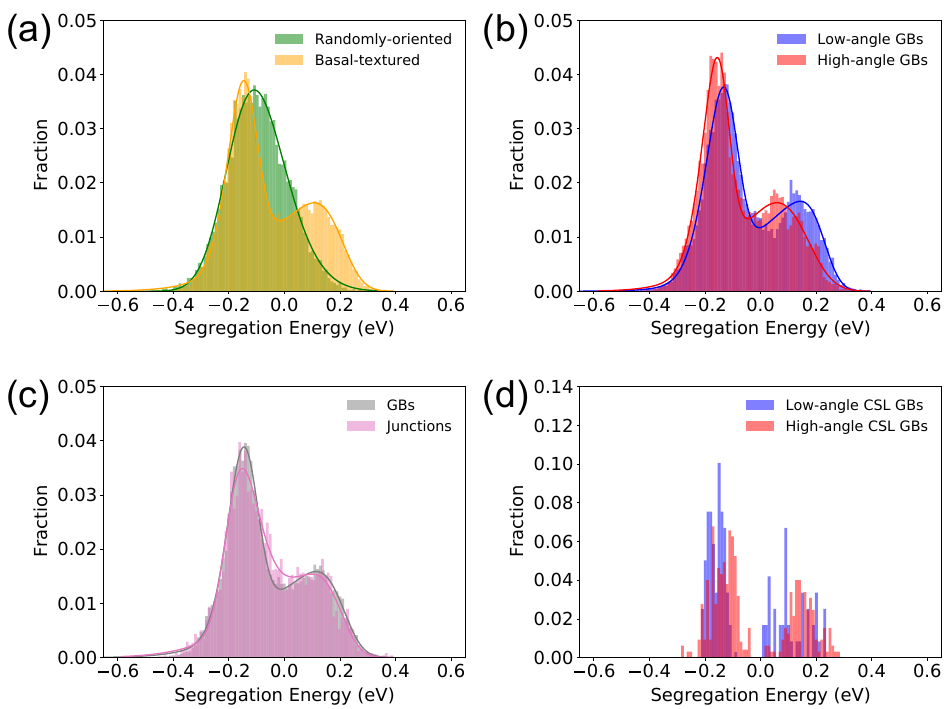}
\caption{Segregation energy spectra of Al solutes at GBs in (a) randomly-oriented polycrystal with $D$=5.8~nm and basal-textured Mg polycrystal with $D$=17.8~nm. The GB segregation energy distribution of the basal-textured Mg polycrystal is further categorized as follows: (b) low-angle ($\theta<15$\textdegree) and high-angle ($\theta>15$\textdegree) GB sites and (c) GB plane sites with only two neighboring grains and GB junction sites with more than two neighboring grains. (d) Segregation energy distribution at low-angle and high-angle \hkl<0001> symmetric tilt CSL GBs in Mg bicrystals. The bin size is 0.01~eV. The segregation energy spectra in polycrystals were fitted to distribution models and are plotted with solid lines.}
\label{fig2}
\end{figure*}

The per-site segregation energies of Al solutes at GBs in basal-textured polycrystals were calculated using molecular statics at 0~K to elucidate the segregation behavior in infinitesimally diluted solid solution alloys. The segregation energy spectrum of the basal-textured polycrystal with $D$=17.8~nm is presented in Figure~\ref{fig2}(a). The data reveals a bimodal distribution, characterized by two distinct peaks located on the negative and positive sides of the distribution, corresponding to favorable and unfavorable segregation, respectively. The negative segregation energy peak is significantly higher and sharper compared to the positive one. In contrast, the energy distribution of the randomly-oriented Mg polycrystal with $D$=5.8~nm, as well as previously reported FCC polycrystalline systems \cite{wagih2019spectrum,wagih2020learning,wagih2022learning,wagih2023can}, follows a skew-normal distribution. This distribution is centered around a single peak, primarily in the negative energy region, indicating that the majority of GB sites are favorable for Al segregation. Notably, the negative peak in the bimodal distribution of the basal-textured polycrystal is slightly more negative than the peak in the skew-normal distribution. Additionally, the basal-textured polycrystal exhibits a higher probability density of unfavorable segregation sites compared to the randomly-oriented polycrystal.

To clarify the contributions of specific local atomic environments to the bimodal distribution of segregation energy in the basal-textured polycrystal, the GB sites were further categorized based on their misorientation angle into low-angle and high-angle GB sites. Additionally, they were classified based on the number of neighboring grains into GB plane sites and GB junction sites. The subspectra of segregation energy in the low-angle and high-angle GB sites are shown in Figure~\ref{fig2}(b), where both exhibit a bimodal distribution similar to the overall spectrum of all GB sites. Notably, both peaks in the bimodal distribution of the high-angle GBs are shifted to the negative side of the spectrum as compared to those of the low-angle GBs, particularly the positive peak. This indicates a greater tendency for Al segregation at high-angle GBs compared to low-angle GBs. Additionally, the negative peak of the high-angle GBs is not only more negative but also higher than the corresponding peak for the low-angle GBs, suggesting a higher probability density for favorable segregation sites in high-angle GBs. 

Figure~\ref{fig2}(c) demonstrates that both GB planes and junctions in the basal-textured Mg polycrystal exhibit similar segregation energy distributions, with a higher density in the negative energy region and a tapering off towards positive energies. The distribution for the GB junctions appears to have a slightly broader spread near the peak region, indicating a more varied range of segregation energies at these sites compared to the GB planes. Additionally, the positive peak in the distribution for the GB junctions is not significant; it appears more as a shoulder rather than a distinct peak, indicating fewer unfavorable segregation sites compared to GB planes.
Similar segregation energy spectra and subspectra of the basal-textured Mg polycrystals were observed in the polycrystals with smaller ($D$=11.3~nm) and larger ($D$=25.2~nm) grain sizes, see Figure S7 in the Supplementary Material. The fitting parameters for the segregation energy distributions are listed in Table SII.

To interpret the bimodal distribution of segregation energy in basal-textured Mg polycrystals containing \hkl<0001> tilt GBs, the per-site segregation energies at 18 \hkl<0001> symmetric tilt CSL GBs in bicrystalline Mg were calculated. The distributions of segregation energy in both low-angle and high-angle symmetric tilt CSL GBs exhibit two distinct peaks, one in the positive region and the other in the negative region, with a gap near zero segregation energy, see Figure~\ref{fig2}(d). Similar to the bimodal distribution in basal-textured Mg polycrystals, the intensity of the negative peak is higher than the positive one, indicating a greater tendency for Al segregation in these regions. Notably, the high-angle CSL GBs exhibit broader peaks in the segregation energy spectrum compared to the low-angle CSL GBs. This suggests that the selected GB sites in the high-angle GBs sample a wider range of segregation energies and, accordingly, local atomic environments. This broader spread reflects the greater complexity and diversity of atomic arrangements at high-angle GBs compared to low-angle GBs. 
 
\begin{figure*}[hbt!]
\centering
\includegraphics[width=\textwidth]{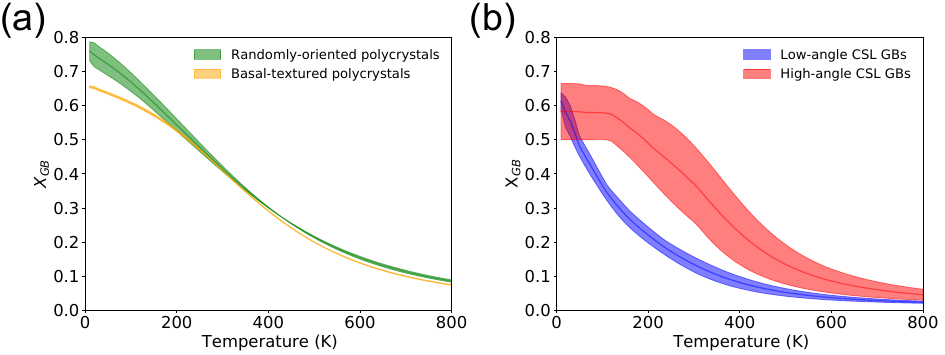}
\caption{GB concentration ($X_{GB}$) of Al solute as a function of temperature at 1\% Al bulk concentration estimated according to the Langmuir–McLean isotherm (Equations~\ref{equ2} and \ref{equ3}) for (a) two randomly-oriented and three basal-textured polycrystals with different grain sizes and (b) eight low-angle ($\Sigma$31, $\Sigma$37, $\Sigma$61, and $\Sigma$73) and ten high-angle ($\Sigma$7, $\Sigma$13, $\Sigma$67, $\Sigma$79, and $\Sigma$97) \hkl<0001> symmetric tilt CSL GBs.}
\label{fig3}
\end{figure*}

Using the Langmuir–McLean isotherm, as introduced in Equations~\ref{equ2} and \ref{equ3}, the solute concentration at individual GBs or within the GB network in dilute solid solutions can be estimated based on given bulk concentrations and temperatures. Figure~\ref{fig3}(a) illustrates how solute concentrations in polycrystals with 1\% Al bulk concentration vary across different temperatures. The three basal-textured polycrystals with different average grain sizes, ranging from 11.3 to 25.2~nm, show a narrow scattering in solute concentration at constant bulk concentration and temperature, suggesting a limited grain size effect on segregation behavior within this length scale. In contrast, the two randomly-oriented polycrystals with grain sizes of 5.8 and 7.2~nm show slightly wider scattering and higher solute concentrations than the basal-textured polycrystals, particularly in the low-temperature regime.

In addition to the differing profiles in segregation energy distribution between the low-angle and high-angle CSL GBs (Figure~\ref{fig2}(d)), the temperature-dependent GB solute concentrations at these CSL GBs with different misorientation angles reveal distinct trends, as presented in Figure~\ref{fig3}(b). The high-angle CSL GBs exhibit significantly wider scattering in solute concentration than the low-angle CSL GBs. Furthermore, the solute concentrations in the high-angle GBs show a plateau before dramatically decreasing above 150~K. In contrast, the solute concentrations at the low-angle CSL GBs drop immediately at finite temperatures. Above 100~K, all high-angle CSL GBs exhibit higher solute concentrations than low-angle CSL GBs, suggesting that high-angle GBs have a greater capacity for Al segregation at elevated temperatures compared to low-angle GBs.

The results in Figures \ref{fig2} and \ref{fig3} highlight the differences in segregation behavior between textured and non-textured polycrystals with various GB characters. This emphasizes the complex intrinsic nature of local atomic environments on solute segregation in Mg polycrystals (see Discussion~\ref{Discuss1}). The next subsection presents the segregation behavior in basal-textured Mg polycrystals within concentrated solid solutions at finite temperatures.

\subsection{\label{Results2}Segregation in concentrated solid solutions at finite temperatures}

\begin{figure*}[hbt!]
\centering
\includegraphics[width=\textwidth]{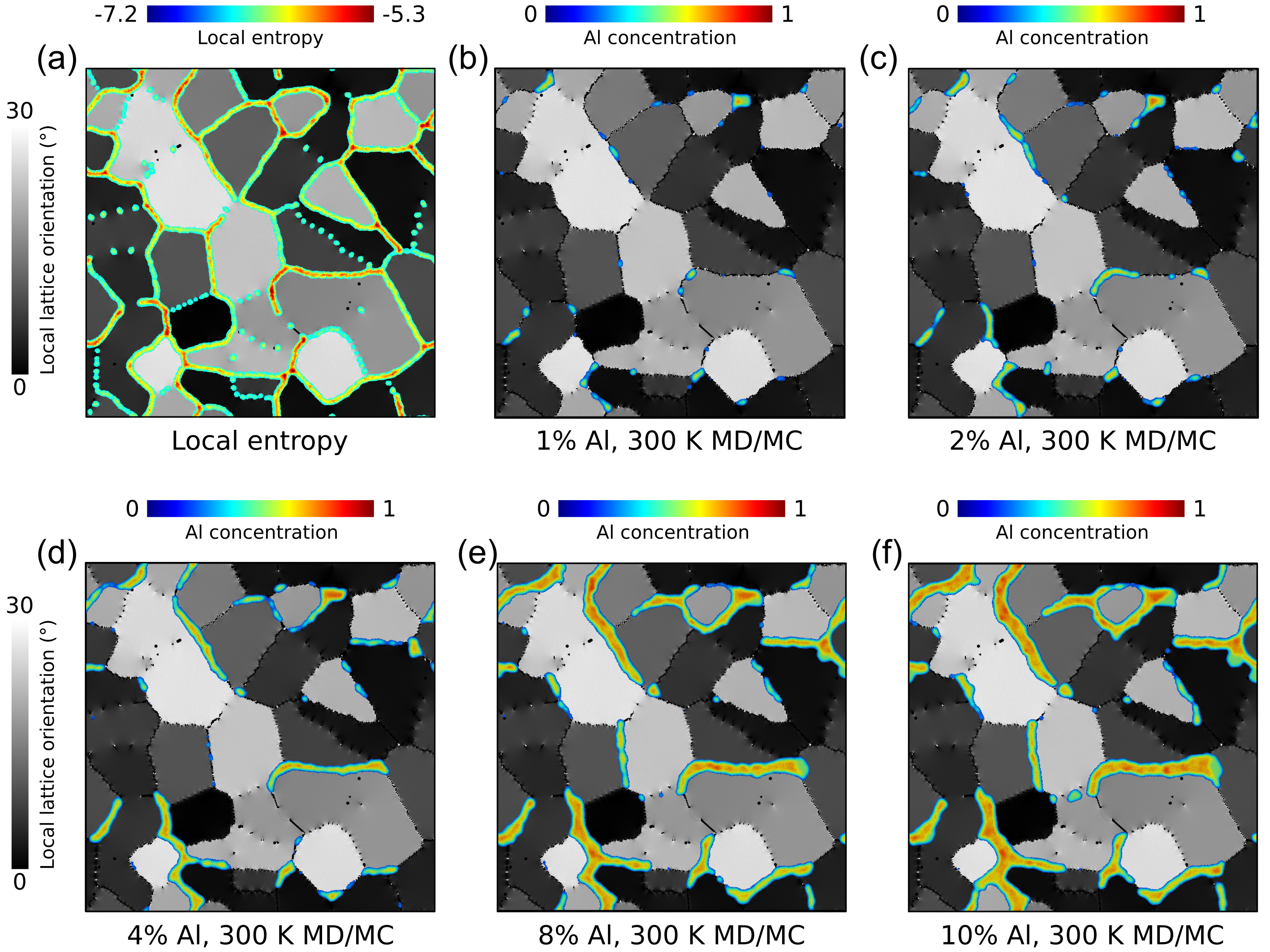}
\caption{(a) Local entropy of the basal-textured polycrystal ($D$=17.8~nm), showing only sites with entropy values $>$-6.5 ($\text{cutoff}=1.6a_{0}$, $\sigma=0.2$). (b-f) Distribution of Al solutes after hybrid MD/MC simulations at 300 K for various Al concentrations ranging from 1\% to 10\% at.\%, displaying only sites with concentrations $>$0.2 at.\%. The local solute concentration and entropy were averaged using a coarse-grain approach, involving the local neighborhood of sites within a cutoff radius of 10~\AA. The background atomistic configuration is colored in greyscale according to the local lattice orientations.}
\label{fig4}
\end{figure*}

The segregation behavior of Al solutes in a concentrated solid solution within a basal-textured polycrystal at finite temperatures was investigated using the hybrid MD/MC approach. The simulations considered the effect of solute concentration at a given temperature on GB phase formation. Figure~\ref{fig4} illustrates the Al solute distribution in a basal-textured polycrystal with $D$=17.8~nm for different target solute concentrations (at.\%) following equilibration at 300~K. At a low solute concentration of 1\% Al, solute enrichment primarily occurs at the triple junctions, as shown in Figure~\ref{fig4}(b). At a higher solute concentration of 2\%, Al continues to segregate to junction sites, with solute enrichment expanding to nearby GB sites (Figure~\ref{fig4}(c)). Interestingly, as the target solute concentration increases, the already enriched regions thicken instead of achieving a homogeneous distribution of Al solutes across other solute-poor GB sites (Figure~\ref{fig4}(d)). Above 8\% solute concentration, the regions of solute enrichment exhibit local Al concentration exceeding 50\% (Figure~\ref{fig4}(e-f)), indicating the formation of Al-rich phases. Notably, solute enrichment, its expansion into neighboring regions, and ultimately culminating in precipitation when the solute concentration reaches a certain threshold, consistently occur at grain boundaries, with no observed intra-grain solute clustering.

\begin{figure*}[htbp!]
\centering
\includegraphics[width=\textwidth]{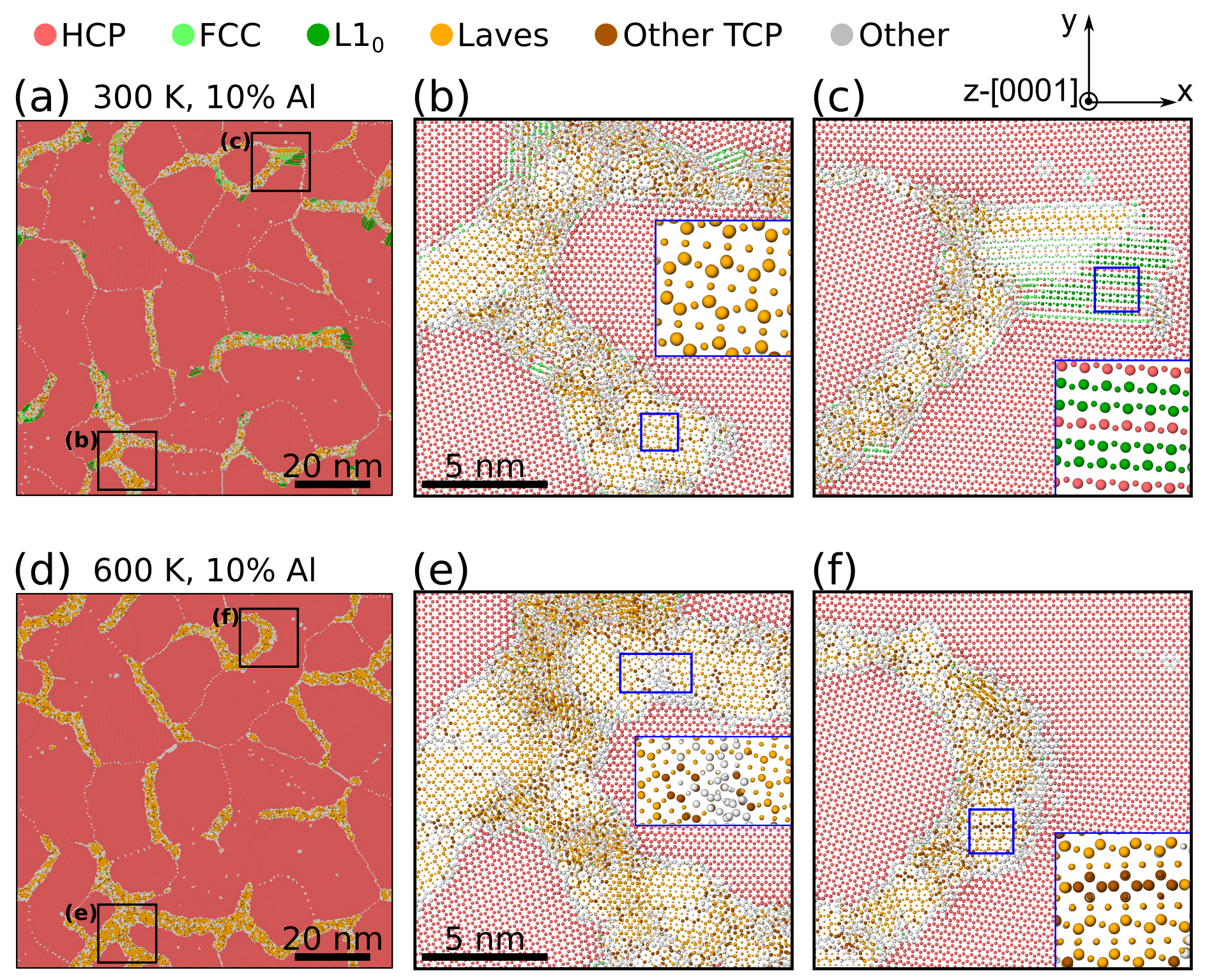}
\caption{Hybrid MD/MC simulations of precipitate formation in basal-textured Mg-10\%Al polycrystal ($D$=17.8~nm) at (a) 300~K and (d) 600~K. (b,c) Zoomed-in views of the regions of interest marked in (a), highlighting the formation of (b) Laves phases and (c) Al-rich FCC and L1\textsubscript{0} phases. (e,f) Zoomed-in views of the regions of interest marked in (d), highlighting the formation of Laves phases and other TCP phases. Atoms are colored according to the PTM and Voronoi indexing algorithms. HCP, FCC and L1\textsubscript{0} atoms were identified using PTM; Laves and other TCP phases were characterized according to Frank-Kasper polyhedra via Voronoi indexing, where Laves phase structures consisting of Z12 and Z16, as well as Z14 and Z15 polyhedra were categorized as Other TCP phase structures. The large and small atoms indicate Mg and Al, respectively.}
\label{fig5}
\end{figure*}

The characteristics of Al-rich phases formed at high solute concentrations were further analyzed by combining the PTM and Voronoi indexing algorithms, which enable the identification of both simple crystal structures with chemical ordering and more complex crystal structures, such as TCP phases. At 10\% Al solute concentration and 300~K, ordered phases, including MgAl\textsubscript{2} Laves phases (i.e., C14, C15, and C36), MgAl L1\textsubscript{0} phase, and Al FCC phase formed at GBs after equilibration (Figure~\ref{fig5}(a)). The columnar Laves phase precipitates were discontinuous and separated by amorphous regions, labeled "Other" in Figure \ref{fig5}. Most Laves phase precipitates were oriented in the $z$-\hkl<2 -1 -1 0> orientation for hexagonal C14 / C36 Laves phases and in the $z$-\hkl<110> orientation for the cubic C15 Laves phase (Figure~\ref{fig5}(b)). The Al FCC phase in the $z$-\hkl<100> orientation formed at GBs with faceted interfaces adjoining the Mg HCP matrix. In contrast, no well-defined interphase boundaries formed between the Laves phase precipitates and the Mg grains (Figure~\ref{fig5}(c)). The MgAl L1\textsubscript{0} phase formed, representing a variation of the solid solution FCC crystal structure with chemical ordering. As shown in the detailed view of the outlined area in Figure~\ref{fig5}(c), coherent twin boundaries formed within the L1\textsubscript{0} phase.  Apart from the formation of the ordered phases, numerous GB structures appeared amorphous or with short-range order containing Frank-Kasper atomic clusters (Z12, Z14, Z15, and Z16), labeled in Figure \ref{fig5} as "Other TCP". 

 Keeping the Al solute concentration at 10\% and raising the temperature to 600~K, precipitation occurred at similar locations in the GB network as it did at 300~K, with predominant MgAl\textsubscript{2} Laves phases observed after equilibration (Figure~\ref{fig5}(d)). Notably, no Al FCC phase was identified. Compared to the discontinuous Laves phases at 300~K, the MgAl\textsubscript{2} Laves phase precipitates at 600~K are much larger and mostly oriented in the $z$-\hkl<2 -1 -1 0> or \hkl<110> directions. Additionally, a smaller fraction of Laves phases oriented in different orientations were observed, introducing internal boundaries within the precipitates (Figure~\ref{fig5}(e)). Other TCP phases, with Zr\textsubscript{4}Al\textsubscript{3} crystal structures consisting of Z14 and Z15 Frank-Kasper atomic clusters, were also observed within the Laves phase precipitates, following the same stacking sequence of the $\mu$ phase (see inset in Figure \ref{fig5}(f)).

\begin{figure*}[hbt!]
\centering
\includegraphics[width=\textwidth]{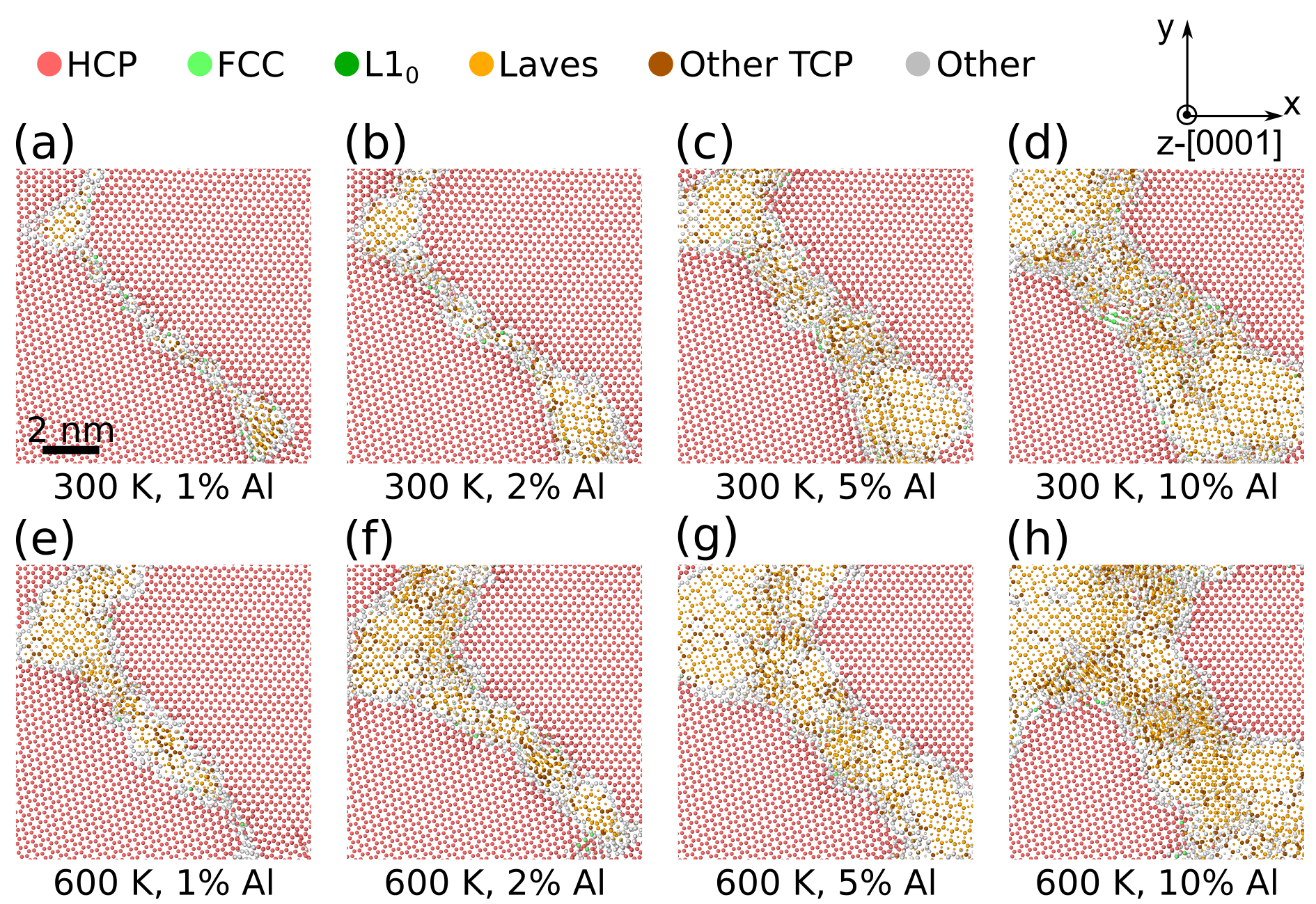}
\caption{Hybrid MD/MC simulations of precipitate formation in basal-textured Mg-Al polycrystal  ($D$=17.8~nm) as a function of different Al solute concentrations. (a-d) 300~K and (e-h) 600~K. The presented region of interest corresponds to the same region (b) and (e) in Figure~\ref{fig5}. Color coding of atoms and phase characterization is the same as in Figure \ref{fig5}.}
\label{fig6}
\end{figure*}

The precipitation of Laves phases at different solute concentrations ($\leq 10\%$ Al) and temperatures (300 and 600~K) is illustrated in a region of interest (cf. Figures \ref{fig5} (b) and (e)) in the basal-textured polycrystal with $D$=17.8~nm, as shown in Figure~\ref{fig6}. At 300~K, when the Al concentration was below 1\%, the GB revealed a few short-range ordered Frank-Kasper atomic clusters, particularly at the junction sites (Figure~\ref{fig6}(a)). At 2\% Al concentration, Laves phase precipitates of a few {\AA}ngström in size formed at those junction sites (Figure~\ref{fig6}(b)). These Laves phase precipitates maintained notable lateral thickening with increasing Al concentration (Figure~\ref{fig6}(c-d)). In contrast to the formation of short-range ordered atomic clusters at 300~K in the low Al concentration range (e.g. 1\%), precipitation of Laves phases was observed even at 1\% solute concentration, when the temperature was raised to 600~K (Figure~\ref{fig6}(e)). These Laves phase precipitates thickened progressively with increasing Al concentration, as shown in Figure~\ref{fig6}(f-h). 

\begin{figure*}[hbt!]
\centering
\includegraphics[width=\textwidth]{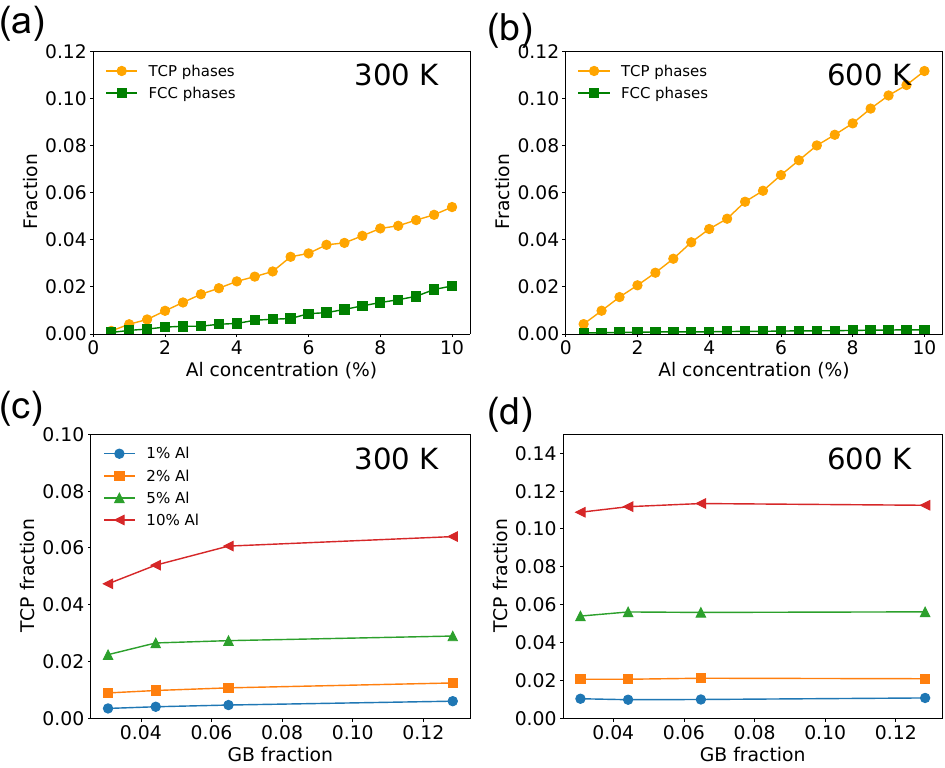}
\caption{Fraction of TCP phases (including Z12, Z14, Z15, and Z16 Frank-Kasper polyhedra characterized via Voronoi indexing) and FCC phases (including FCC and L1\textsubscript{0} characterized via PTM) as a function of target Al concentration in the basal-textured polycrystal ($D$=17.8~nm) after the hybrid MD/MC simulations at (a) 300~K and (b) 600~K. Fraction of TCP phases as a function of GB fraction in four basal-textured polycrystals with different grain sizes ($D$=25.2, 17.8, 11.3, and 5.6~nm) at (c) 300~K and (d) 600~K.}
\label{fig7}
\end{figure*}

The statistics of the number of atoms in TCP phases (including Laves phases and other TCP phases) and FCC phases (including FCC and L1\textsubscript{0} phases) at different Al concentrations and temperatures in the basal-textured polycrystal with $D$=17.8~nm are shown in Figure~\ref{fig7}(a-b). 
The evolution of the TCP phases at 300 and 600~K exhibits a linear increase with increasing solute concentration (Figure~\ref{fig7}(a-b)). The fraction of TCP phases reaches 11\% at 600~K, which is roughly two times the amount observed at 300~K. Similar to the observed linear increase of TCP phase at 300~K, the fraction of FCC phases steadily increases with increasing solute concentration, as shown in Figure~\ref{fig7}(a). This increase follows a non-linear trend, with the rate of increase becoming more pronounced as the Al concentration rises. Below 10\% Al concentration, TCP phases dominate the GB regions, with their fraction being higher than that of FCC phases. Conversely, at 600~K, the fraction of FCC phases remains approximately 0\%, correlating well with the observation in Figure~\ref{fig6}(e-h).

The hybrid MD/MC simulations were also performed on basal-textured polycrystals with varying grain sizes, i.e. GB fractions for different Al solute concentrations and equilibrium temperatures. At 300~K, the number of atoms in TCP phases increases with increasing GB fraction (or decreasing grain size), particularly at low GB fractions, as sample dimensions and total number of atoms remain roughly constant across polycrystals with different grain sizes (Figure~\ref{fig7}(c)). This increase is more pronounced for higher solute concentrations. Interestingly, almost no grain size effect on the TCP phase fraction was observed at 600~K across all solute concentration environments (Figure~\ref{fig7}(d)). The types of GB phases remain consistent at different grain sizes at both 300~K and 600~K, with only TCP phases formed at 600~K, regardless of grain size. The interpretation of the hybrid MD/MC observations on temperature and solute concentration-dependent GB phase formation is discussed in section Discussion~\ref{Discuss2}.

\section{\label{Discuss}Discussion}

\subsection{\label{Discuss1}Segregation energy and local atomic environment spectra}

\begin{figure*}[hbt!]
\centering
\includegraphics[width=\textwidth]{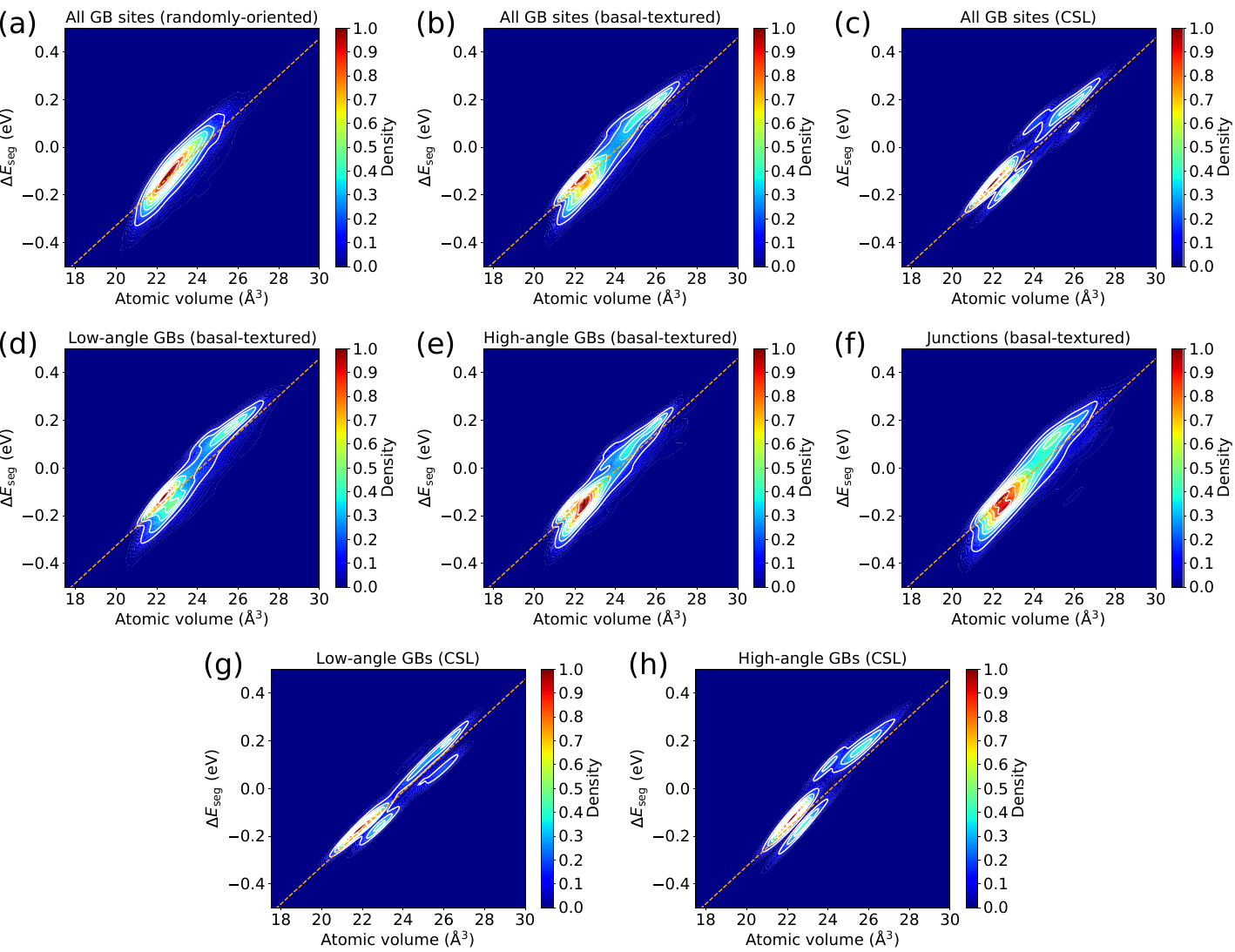}
\caption{Probability density contour plots of the atomic volume ($V_{GB}^{i}$) and segregation energy ($\Delta E_{seg}^{i}$) of GB sites in the (a) randomly-oriented polycrystal ($D$=5.8~nm), (b) basal-textured polycrystal ($D$=17.8~nm), and (c) \hkl<0001> symmetric tilt CSL GBs. (d) Low-angle GBs, (e) high-angle GBs, and (f) GB junctions in the basal-textured polycrystal. (g) Low-angle GBs and (h) high-angle GBs in the tilt CSL GBs. The orange dashed lines indicate the linear elastic model \cite{huber2014atomistic,huber2018machine}: $\Delta E_{seg}^{i} = B ( V_{bulk}^{Al} / V_{bulk} - 1)(V_{GB}^{i}-V_{bulk}^{i}) + \Delta E_{bond}$, where $B$ is the bulk modulus of Mg, $V_{bulk}$ is the volume of a bulk Mg, $V_{bulk}^{Al}$ is the volume of the bulk Mg containing a Al solute, and $V_{bulk}^{i}$ is the Mg atomic volume. $\Delta E_{bond}$ is a fitting parameter that accounts for the energy contribution arising from changes in the bonding environment. The best fit for the $\Delta E_{bond}$ value is -0.11~eV.}
\label{fig8}
\end{figure*}

Textured polycrystalline structures represent an important class of technical materials that exhibit significant anisotropy in their mechanical response and have been widely reported in various metallic systems \cite{thompson1995texture,zhang2022atomistic}. The presence of a specific grain orientation distribution within these polycrystalline structures leads to a unique variety of tilt GBs, sampling only a subset of the full range of the local atomic environment space found in randomly-oriented polycrystalline structures. In the current work, the studied basal-textured Mg polycrystal showed a bimodal distribution of the segregation energy profile, predicted by a White-Coghlan extension of the McLean-style approach. This bimodal distribution was also evident when the GB network of the $z$-\hkl[0001] polycrystal was broken down into low-angle and high-angle boundaries. For the special case of symmetric tilt CSL structures amongst these boundaries, the segregation spectra of both low-angle and high-angle \hkl<0001> CSL GBs revealed two distinct peaks associated with negative and positive segregation energies, respectively (Results~\ref{Results1}). This behavior suggests a clear separation in the local atomic environment space of GB sites in the tilt CSL GBs. Huber et al. \cite{huber2014atomistic,huber2018machine} developed a linear elastic model that effectively correlates the atomic volume at GBs to the per-site segregation energy in highly symmetric GBs. Figure~\ref{fig8} shows a good linear correlation between atomic volume and segregation energy in all simulated cases of randomly-oriented and basal-textured polycrystals, as well as CSL GBs. As shown, GB sites with larger atomic volumes (extensive sites) exhibit a less favorable segregation tendency, and vice versa. This is because the Al atoms, which have a smaller atomic radius than Mg, are more likely to segregate to compressive sites that have smaller atomic volumes compared to bulk Mg sites. The randomly-oriented polycrystal exhibits one hot spot of high density located in the negative segregation energy region, see Figure~\ref{fig8}(a). In contrast, the basal-textured polycrystal shows two distinct high-density spots in both the negative and positive segregation energy regions (Figure~\ref{fig8}(b)), corresponding to the compressive and extensive GB site regions, respectively.

The GB network in the basal-textured polycrystal displays a CSL-like feature, characterized by repeated compressive and extensive GB sites along dislocation arrays or structural units, as shown in Figure~\ref{fig1}(c). However, compared to the CSL GBs, the basal-textured polycrystal exhibits a significantly broader distribution in the density contour plot, demonstrating the limited sampling of the local atomic environment space by the CSL tilt GB sites when asymmetric tilt GBs and GB junctions are excluded (Figure~\ref{fig8}(c)). The low-angle GBs in the polycrystal exhibit two high-density spots in both the negative and positive segregation energy regions, though not as pronounced as in the CSL GBs. For the high-angle GBs, the density in the positive segregation energy region becomes less pronounced compared to the low-angle GBs, while the density is more spread out on the negative side. The GB junctions consist of a more varied range of local atomic environments with favorable segregation sites. This is evidenced by a broader spread of the high-density spot compared to the low-angle and high-angle GBs, see Figure~\ref{fig8}(f). Additionally, the GB junctions exhibit a smoother transition from the highly concentrated negative segregation energy region to the less concentrated positive region, representing the diminished positive peak in the segregation energy subspectra (Figure~\ref{fig2}(c)). The linear elastic model fails to account for the solute segregation spectrum and the variations in the local atomic environment, particularly in polycrystals. This is because the atomic volume only characterizes the excess free volume within the nearest neighboring cutoff, whereas the local structural rearrangement after introducing a substitutional solute atom leads to a structural relaxation that extends beyond the nearest neighbors \cite{pei2023atomistic}. 

The GB junctions are the most favorable locations for solute enrichment after equilibrium at finite temperatures, as demonstrated by the hybrid MD/MC simulations, see Figure~\ref{fig4}. The per-site local entropy was used to characterize the ordering of the local atomic environments in the GB region of the basal-textured polycrystal. A higher local entropy value indicates a more disordered local structure. A strong correlation was found between regions with high local entropy and regions of high solute enrichment in the MD/MC simulations (Figure~\ref{fig4}). This indicates that disordering assists GB structural transition and that the disordered GB regions act as nucleation sites for GB phase formation. This correlation highlights the importance of understanding local atomic environments in predicting solute segregation and its effects on material properties.

\subsection{\label{Discuss2}Grain boundary phase formation}

\begin{figure}[hbt!]
\centering
\includegraphics[width=0.5\textwidth]{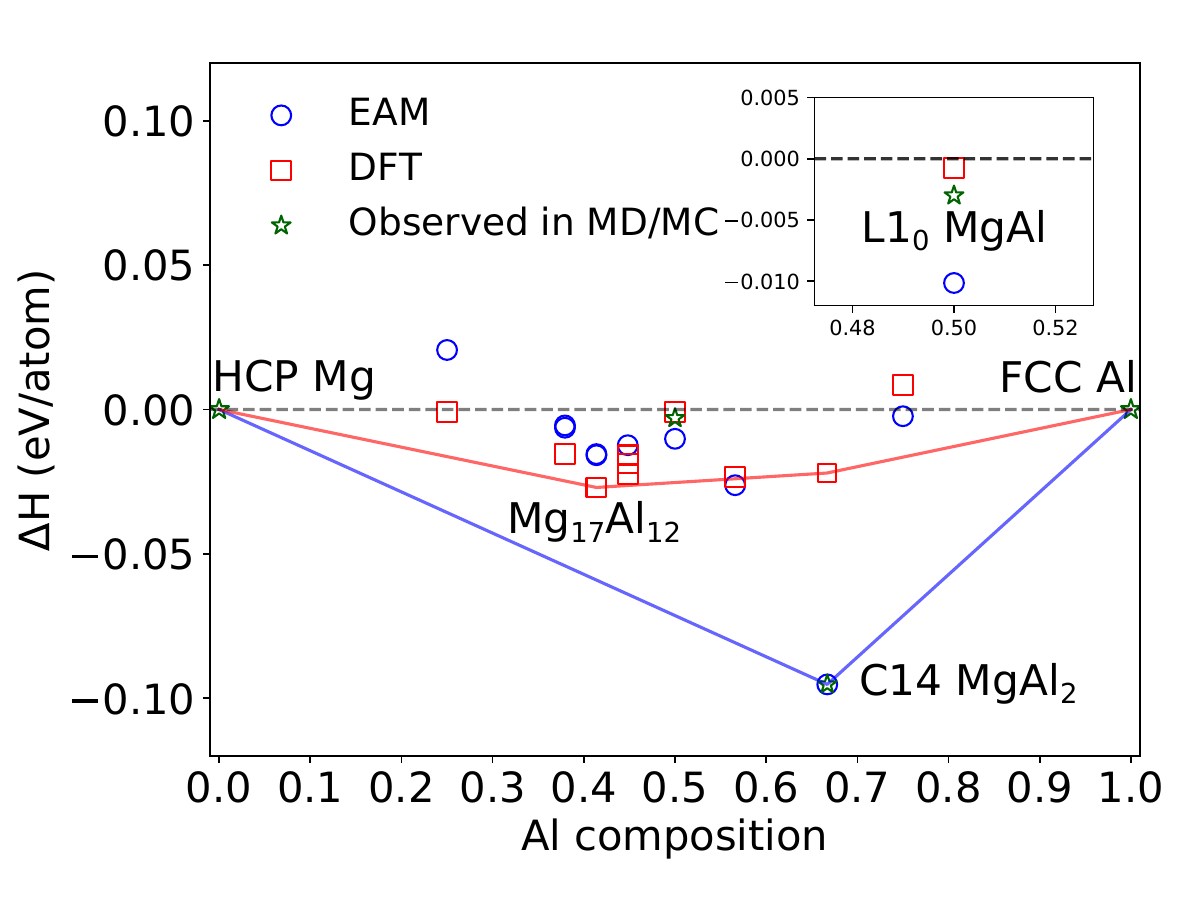}
\caption{Enthalpy of formation of different phases in the Mg-Al  thus material properties, can be
tailored by adjusting the chemicalsystem from the EAM and DFT calculations. The observed phases in the hybrid MD/MC simulations were highlighted with star symbols.}
\label{fig9}
\end{figure}

At finite temperatures, the MgAl\textsubscript{2} Laves phase was demonstrated in Results~\ref{Results2} as the predominant secondary phase to precipitate at the GB, particularly at higher temperatures. Conversely, at room temperature, the Laves phase co-exists with FCC phases, including Al FCC and MgAl L1\textsubscript{0} phases. 
The preference for the formation of the MgAl\textsubscript{2} Laves phase in this study can be rationalized by its low enthalpy of formation, as predicted by the EAM potential, see Figure~\ref{fig9}. Interestingly, the formation of Laves phases at GBs varies significantly with changes in temperature and chemical composition, as shown in Figure~\ref{fig7}(a-b). 
At 300~K, the Laves phase precipitates appeared relatively small and discontinuous separated by other competitive GB phases and amorphous regions even at 10\% Al concentration. As the GB fraction increased with decreasing grain size, the fraction of Laves phases was observed to rise, particularly at low GB fractions, under the same thermodynamic variables. This is due to the higher fraction of GB sites favorable for Laves phase nucleation and growth within the polycrystals at 300~K. 

At 600~K, the Laves phase precipitates were continuous and much larger than those at 300~K. In addition, the Laves phase fraction remained almost consistent across different GB fractions at 600~K, indicating that the energy barriers for heterogeneous nucleation of precipitates and their growth into the intragranular regions were overcome with the assistance of thermal fluctuations. No general or intragranular precipitation was observed in this study because the MD/MC approach only provides a thermodynamic equilibrium state without modeling the dynamic processes of solid solution decomposition and precipitation. As a result, the simulations consistently show GB precipitation to be energetically more favorable than intragranular precipitation. This leads to virtually no solute clustering inside the grains in the simulated polycrystal.

Among the energy contributions of Gibbs free energy change to secondary phase formation, interphase boundary energy and strain energy due to lattice mismatches also serve as barriers to nucleation. In this study, the majority of Laves phase precipitates were oriented in the $z$-\hkl<2 -1 -1 0> or \hkl<110> orientation in the basal-textured ($z$-\hkl<0001>) Mg polycrystals (see Figure~\ref{fig5}(a,d)). 
The most energetically favorable orientation relationship between the Laves phase precipitate and the Mg matrix is \hkl<0001>\textsubscript{C14,C36} $\parallel$ \hkl<0001>\textsubscript{Mg} or \hkl<111>\textsubscript{C15} $\parallel$ \hkl<0001>\textsubscript{Mg} as widely reported in Mg alloys \cite{suzuki2005solidification,zubair2023laves,guenole2021exploring}, indicating minimized interface energy and strain energy with this orientation relationship. 
The formation of Laves phases in the less favorable orientations observed here can be attributed to the limited dimension of the polycrystals in the $z$-direction (3.13~nm). For the C14 MgAl\textsubscript{2} Laves phase, the equilibrium lattice constants are $a_{0}=5.37$~\AA\ and $c_{0}=8.84$~\AA. The minimal mismatch strains of the Mg matrix in the $z$-\hkl<0001> direction with the columnar Laves phase in $z$-\hkl<0001> and $z$-\hkl<2 -1 -1 0> directions are 13.8\% and 3.7\%, respectively. The orientation relationship of \hkl<2 -1 -1 0>\textsubscript{C14} $\parallel$ \hkl<0001>\textsubscript{Mg} exhibits a much lower misfit strain energy and presumably leads to an overall lower energy barrier.

In addition to the formation of Laves phases, other metastable intermetallic phases were stabilized at GBs with specific thermodynamic variables, e.g., the MgAl L1\textsubscript{0} phase at 300~K and the Mg\textsubscript{4}Al\textsubscript{3} TCP phase at 600~K, which can be categorized as defect phases \cite{korte2022defect}. While the hybrid MD/MC approach primarily determines GB phases thermodynamically stable, it does not provide information on the kinetics of GB phase nucleation and growth. However, the approach does yield valuable insights into the GB segregation behavior of textured polycrystals under thermodynamic equilibrium states across a wide range of chemical potential space. By adjusting the chemical potential, it becomes feasible to tailor GB structures and consequently material properties along the reaction pathway of GB phase formation, which entails managing multiple energy barriers. For instance, experimental observations in Mg-Zn alloys, have shown that short-range order structures, such as Frank-Kasper atomic clusters near defects \cite{yang2018precipitation} can be stabilized within a specific chemical potential window by the nucleation barrier of TCP phases.

The EAM potential favored the formation of MgAl\textsubscript{2} Laves phase, as evidenced by its overwhelmingly low formation enthalpy compared to other intermetallic compounds. This resulted in other intermetallics positioned above the convex-hull being thermodynamically metastable or unstable. This differs from the DFT results, where the MgAl\textsubscript{2} Laves phase lies on the convex-hull, and Mg\textsubscript{17}Al\textsubscript{12} emerges as the intermetallic phase with the lowest formation enthalpy (Figure~\ref{fig9}). 
It is worth mentioning that while there are discrepancies due to artifacts in the interatomic potential, there are also similarities between the simulation and the experimentally observed GB phase formation of the Mg-Al material system. For instance, cellular precipitates of the Mg\textsubscript{17}Al\textsubscript{12} phase form discontinuously at high-angle and disordered GBs in Mg-Al alloys after aging in the low-temperature regime, as reported in previous experiments \cite{clark1968age,braszczynska2009discontinuous}. In contrast, continuous Mg\textsubscript{17}Al\textsubscript{12} precipitates within grains have been observed after aging at a high temperature of 623~K \cite{braszczynska2009discontinuous}. This experimentally observed temperature-dependent continuity of GB phase formation is similar to the MC/MD outcomes in this work.
In general, the approach utilized in the current study to unravel the atomistic origins of GB phase formation in the chemical potential space, can be applied to other relevant material systems, such as Mg-Zn, Mg-Cu, and Mg-Ni alloys where Laves phase prototypes (C14 MgZn\textsubscript{2}, C15 MgCu\textsubscript{2}, and C36 MgNi\textsubscript{2}) exhibit the lowest formation enthalpy. 

\subsection{\label{Discuss3}Limitations of segregation and simulation models}

\begin{figure}[t]
\centering
\includegraphics[width=1.0\columnwidth]{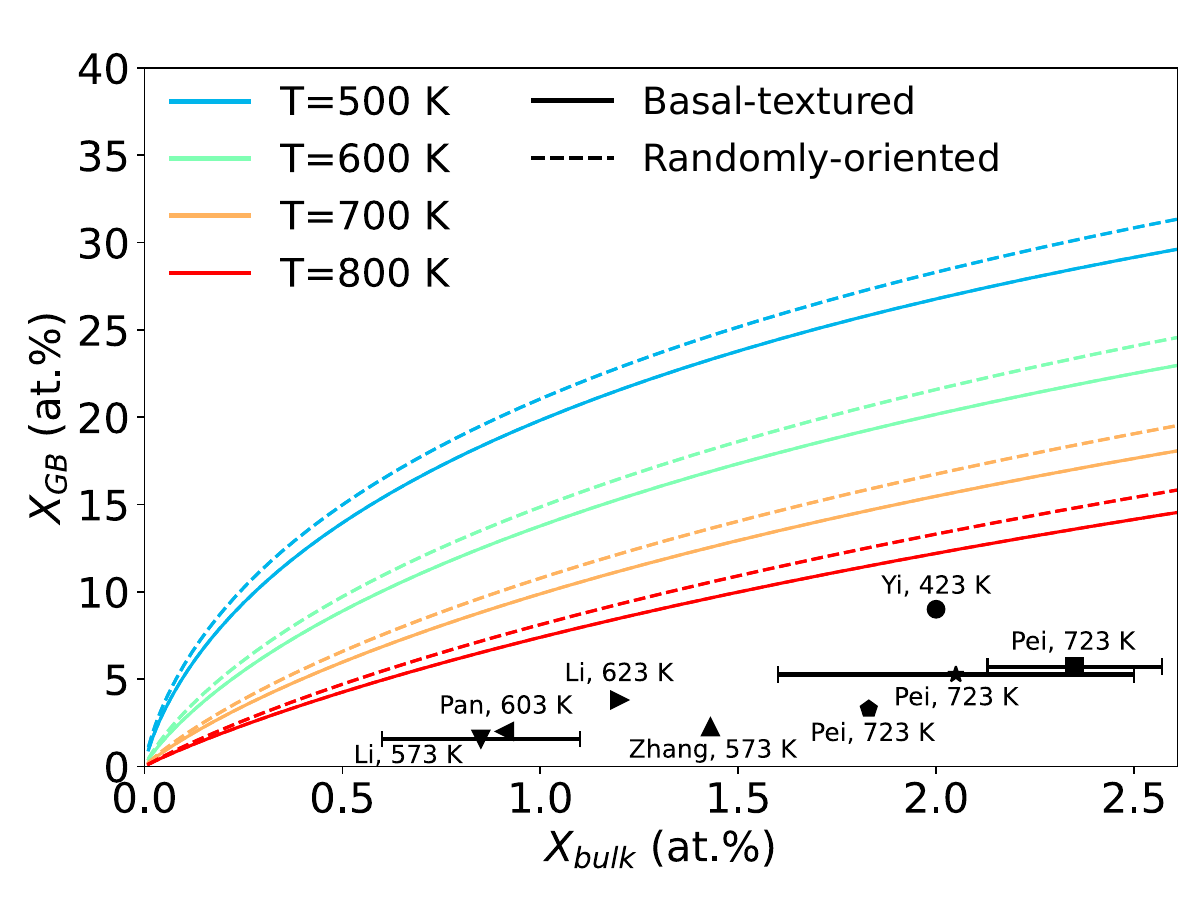}
\caption{The concentrations of Al solutes at GBs within the simulated Mg basal-textured and randomly-oriented polycrystals as a function of their bulk solute concentrations across various temperature regimes following the Langmuir–McLean isotherm, as denoted in Equations~\ref{equ2} and~\ref{equ3}. Experimental data \cite{pei2021grain,li2022elucidation,li2022elucidating,pei2022synergistic,yi2023interplay,zhang2023anisotropic,pan2020mechanistic,pei2022effect} are illustrated as black dots on the corresponding graphs, labeled with the first authors' names and the thermomechanical processing temperatures of the Mg alloys.}
\label{fig10}
\end{figure}

The predicted $X_\text{GB}$ values according to the Langmuir–McLean isotherm (Equations~\ref{equ2} and~\ref{equ3}) were compared to experimentally measured local chemical distributions obtained at thermomechanical temperatures prior to quenching and subsequent sample preparation for microscopic and chemical characterization (Figure~\ref{fig10}). The experimental values were measured using APT or energy-dispersive X-ray spectroscopy (EDS). An overestimation of $X_\text{GB}$ using the Langmuir–McLean isotherm, which employs per-site segregation energy at 0 K, was identified. This could be primarily attributed to ignoring solute-solute interactions for concentrated solid solutions and the vibrational entropy contribution for the high-temperature regime. The contribution of solute-solute interactions to non-linearity in segregation behavior can be accounted for by incorporating the solute interaction term in the Fowler-Guggenheim isotherm~\cite{fowler1939statistical}. The vibrational entropy contribution to the segregation free energy can be estimated through thermodynamic integration \cite{menon2024atomistic} or harmonic approximation \cite{tuchinda2023vibrational}. Apart from the limitations in the segregation isotherm, the limitations of interatomic potential in accurately capturing the variation in solute segregation behaviors across the diverse local atomic environments of GBs have been reported \cite{wagih2022learning}. Hybrid multiscale modeling approaches, such as quantum mechanical/molecular mechanical (QM/MM), could offer QM-level accurate results while circumventing the size constraints typically associated with ab-initio calculations.

As demonstrated by the segregation energy spectrum and solute enrichment at specific GBs and junctions in this work, the segregation behavior of one GB cannot represent the entire GB network due to incomplete coverage of the local atomic environment space. However, experimental characterization is often performed on selected GBs. As revealed in our recent APT study~\cite{pei2023atomistic}, significant variations have been observed in solute concentrations at GBs with distinct macroscopic characteristics. Furthermore, Mg-Al alloy systems often contain additional solute elements beyond Al in experiments \cite{pei2021grain,li2022elucidation,li2022elucidating,pei2022synergistic,zhang2023anisotropic,pan2020mechanistic,pei2022effect}, necessitating the consideration of synergistic effects \cite{mouhib2024exploring}. For the validation of segregation and simulation models, more systematic experimental characterization of solute concentration at various GBs and junctions across a wide range of chemical potential space is required.

\FloatBarrier

\section{Conclusions}

In this work, we investigated the GB segregation behavior of solute atoms in dilute and concentrated solid solution Mg-Al alloys at different temperatures and chemical compositions using atomistic simulations. From the obtained results, the following conclusions can be inferred: 

\begin{itemize}
\item The segregation energy spectra of basal-textured Mg polycrystals exhibit a distinct bimodal distribution, in contrast to the skew-normal distribution observed in randomly-oriented polycrystals. 

\item High-angle \hkl<0001> tilt GBs exhibit a greater tendency for solute enrichment than low-angle \hkl<0001> tilt GBs, but solute enrichment primarily occurs at the GB junctions. It originates from the wider range of local atomic environments accessible and the increased local structural entropy, offering more sites with favorable segregation energies.

\item The formation of Al-rich secondary phases, including  MgAl\textsubscript{2} Laves phases, Al FCC phases, and other metastable intermetallic phases such as MgAl L1\textsubscript{0} and Mg\textsubscript{4}Al\textsubscript{3} TCP phases, was identified at GBs.

\item GB phase formation is strongly dependent on the chemical composition and temperature. At 300~K, Laves phase precipitates are discontinuous separated by other competitive GB phases and amorphous regions. At 600~K, Laves phase precipitates become continuous, and the nucleation and growth barriers are surpassed with the assistance of thermal fluctuations.

\item MD/MC simulations demonstrate a strong correlation between high local entropy regions and elevated solute enrichment, highlighting disordered GB regions as effective nucleation sites for GB phase formation.

\item GB structures, and thus material properties, can be tailored by adjusting the chemical potential along the reaction pathway of GB phase formation, which involves multiple energy barriers.
\end{itemize}

\section*{Acknowledgments}
Z.X. and T.A.S. acknowledge the financial support by the DFG 
(Grant Nr. 505716422). Z.X. and S.K.K. acknowledge the financial support by the DFG through the SFB1394 Structural and Chemical Atomic Complexity – From Defect Phase Diagrams to Material Properties, project ID 409476157. Additionally, Z.X. and S.K.K. are grateful for funding from the European Research Council (ERC) under the European Union’s Horizon 2020 research and innovation program (grant agreement No. 852096 FunBlocks). J.G. acknowledges funding from the French National Research Agency (ANR), Grant ANR-21-CE08-0001 (ATOUUM) and ANR-22-CE92-0058-01 (SILA). Simulations were performed with computing resources granted by RWTH Aachen University under project (p0020267). 


\bibliographystyle{elsarticle-num}
\bibliography{references}

\clearpage

\onecolumngrid
\begin{center}
\Large Supplementary Material
\end{center}

\setcounter{figure}{0}
\setcounter{table}{0}
\renewcommand{\figurename}{Figure S}
\renewcommand{\tablename}{Table S}

\begin{figure*}[hpt!]
\centering
\includegraphics[width=\textwidth]{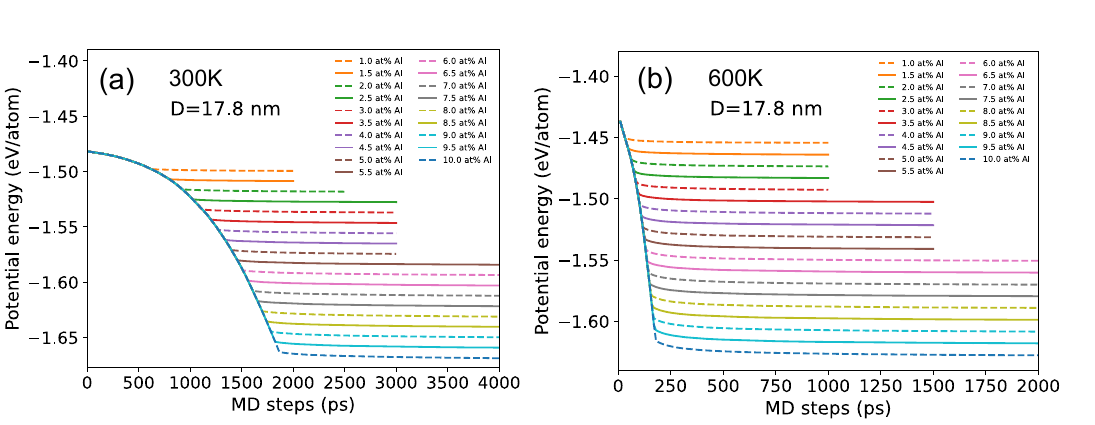}
\caption{Evolution of per-atom potential energy during the hybrid MC/MD simulations in basal-textured polycrystals with grain size ($D$)=17.8 nm and different target concentrations at (a) 300 K and (b) 600 K.}
\label{figS1}
\end{figure*}

\begin{figure*}[hpt!]
\centering
\includegraphics[width=\textwidth]{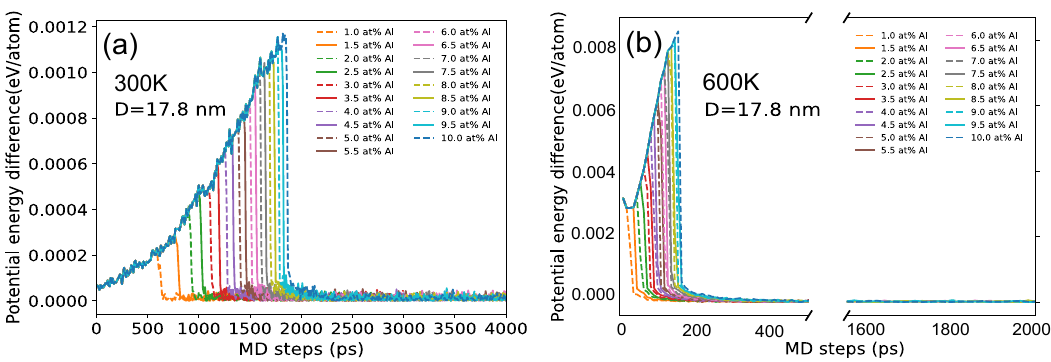}
\caption{Evolution of difference in per-atom potential energy between each step during the hybrid MC/MD simulations in basal-textured polycrystals with grain size ($D$)= 17.8 nm and different solute concentrations at (a) 300 K and (b) 600 K.}
\label{figS2}
\end{figure*}

\begin{figure*}[hpt!]
\centering
\includegraphics[width=\textwidth]{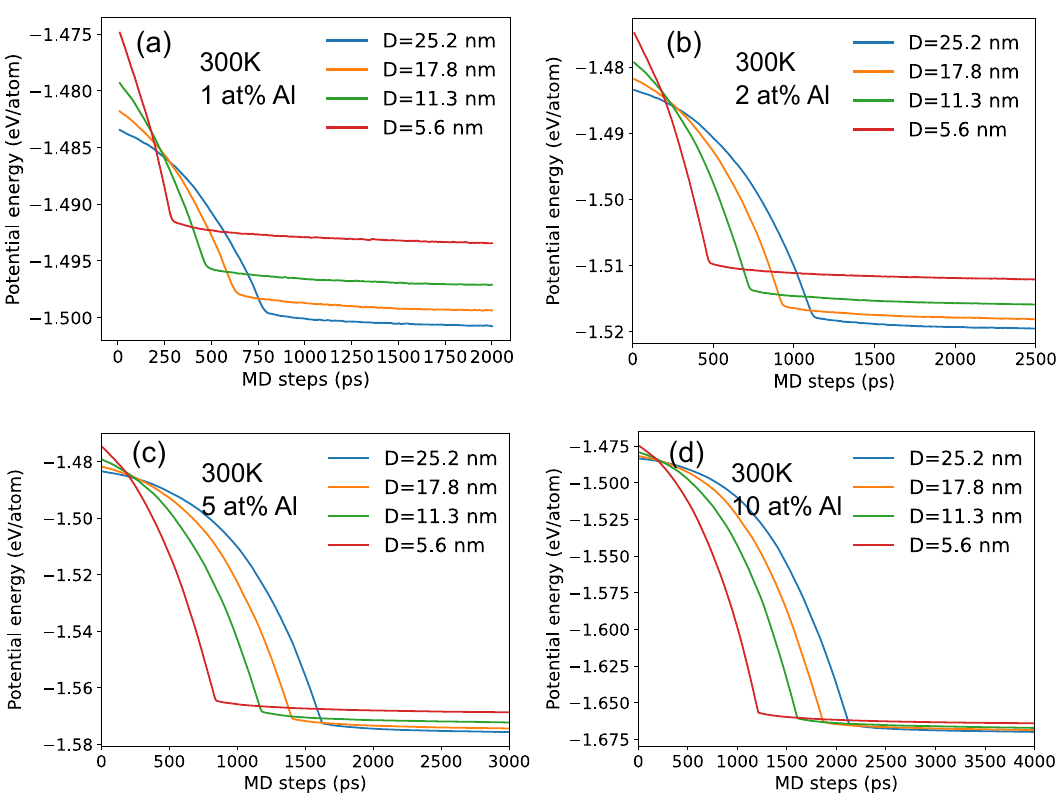}
\caption{Evolution of per-atom potential energy during the hybrid MC/MD simulations at 300 K in basal-textured polycrystals with different solute concentrations of (a) 1 at$\%$ Al, (b) 2 at$\%$ Al, (c) 5 at$\%$ Al and (d) 10 at$\%$ Al.}
\label{figS3}
\end{figure*}

\begin{figure*}[hpt!]
\centering
\includegraphics[width=\textwidth]{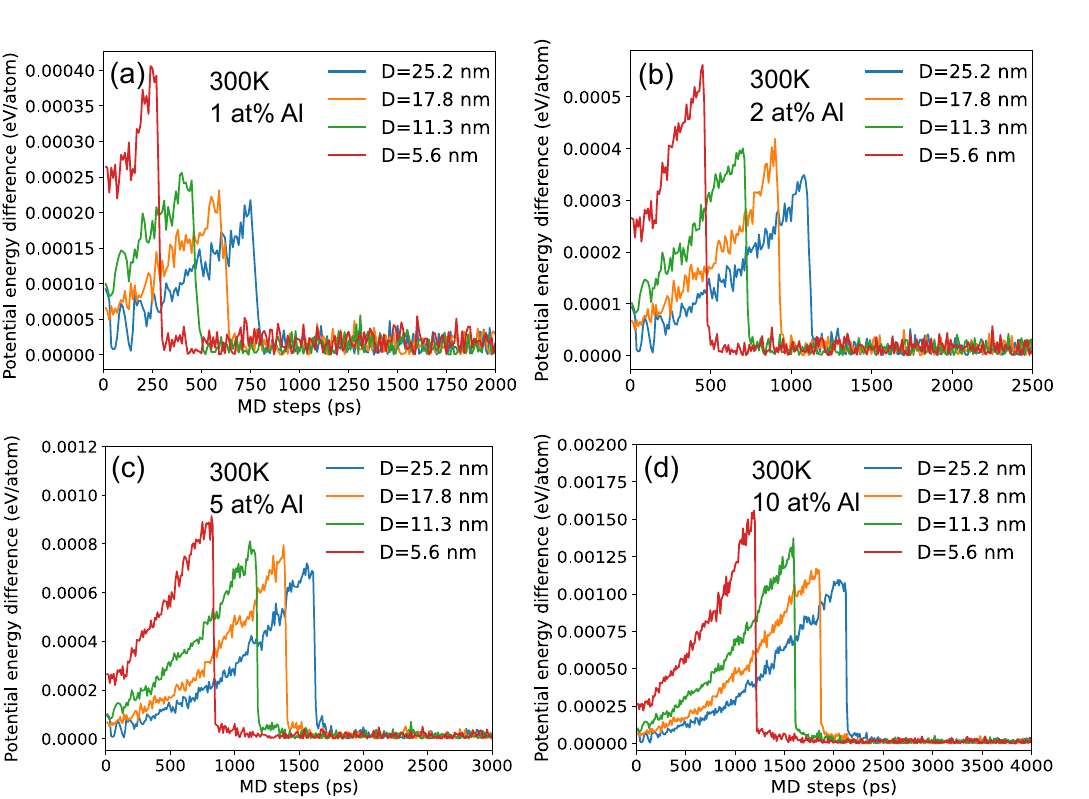}
\caption{Evolution of difference in per-atom potential energy between steps during the hybrid MC/MD simulations at 300 K in basal-textured polycrystals with different solute concentrations of (a) 1 at$\%$ Al, (b) 2 at$\%$ Al, (c) 5 at$\%$ Al and (d) 10 at$\%$ Al.}
\label{figS4}
\end{figure*}

\begin{figure*}[hpt!]
\centering
\includegraphics[width=\textwidth]{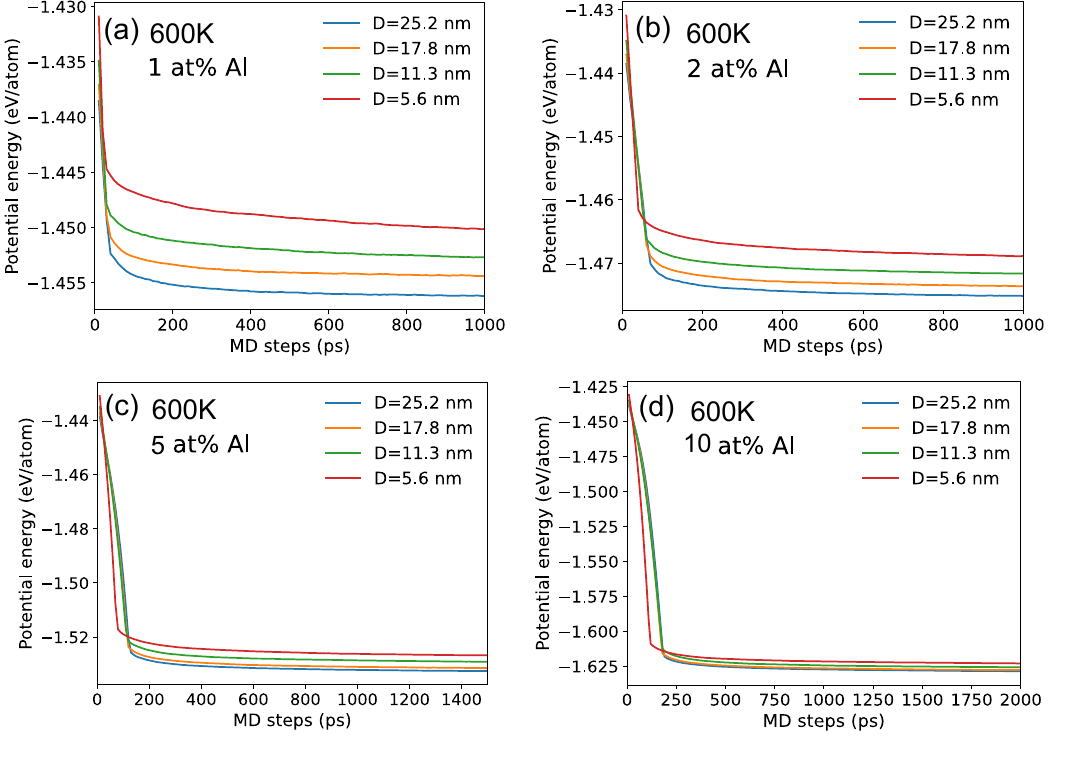}
\caption{Evolution of per-atom potential energy during the hybrid MC/MD simulations at 600 K in basal-textured polycrystals with different solute concentrations of (a) 1 at$\%$ Al, (b) 2 at$\%$ Al, (c) 5 at$\%$ Al and (d) 10 at$\%$ Al.}
\label{figS5}
\end{figure*}

\begin{figure*}[hpt!]
\centering
\includegraphics[width=\textwidth]{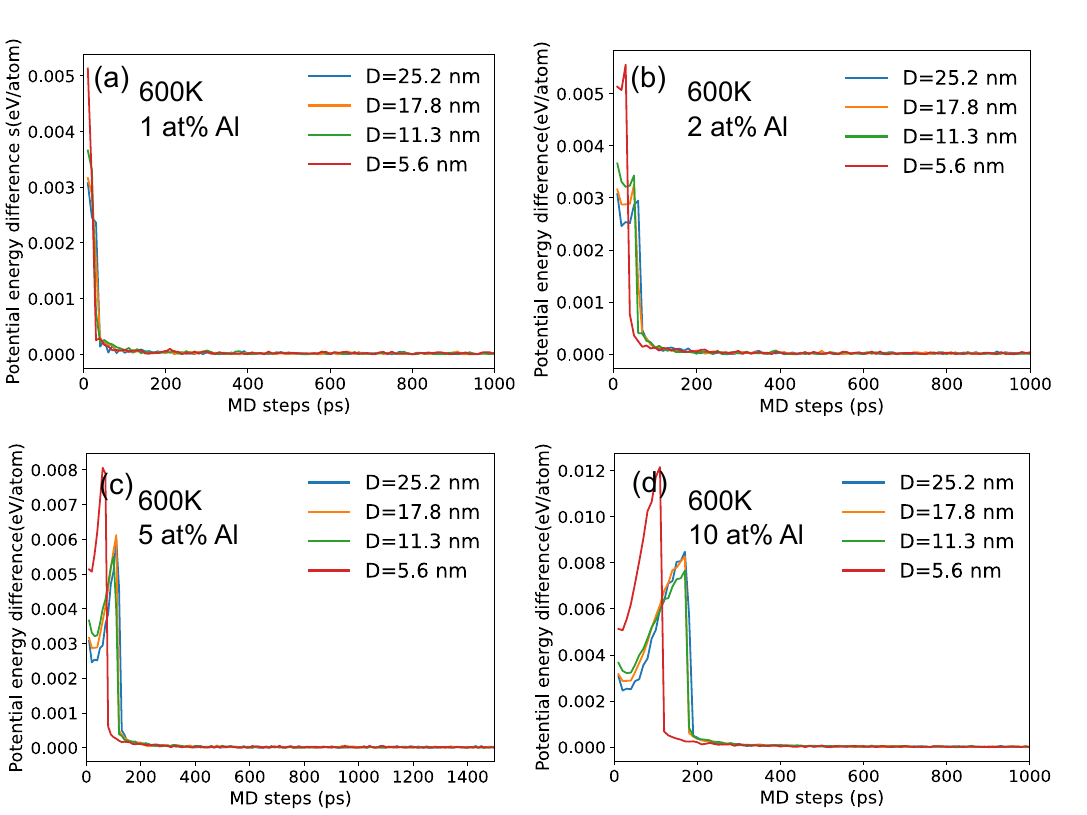}
\caption{Evolution of difference in per-atom potential energy between steps during the hybrid MC/MD simulations at 600 K in basal-textured polycrystals with different solute concentrations of (a) 1 at$\%$ Al, (b) 2 at$\%$ Al, (c) 5 at$\%$ Al and (d) 10 at$\%$ Al.}
\label{figS6}
\end{figure*}

\FloatBarrier

\begin{figure*}[hpt!]
\centering
\includegraphics[width=\textwidth]{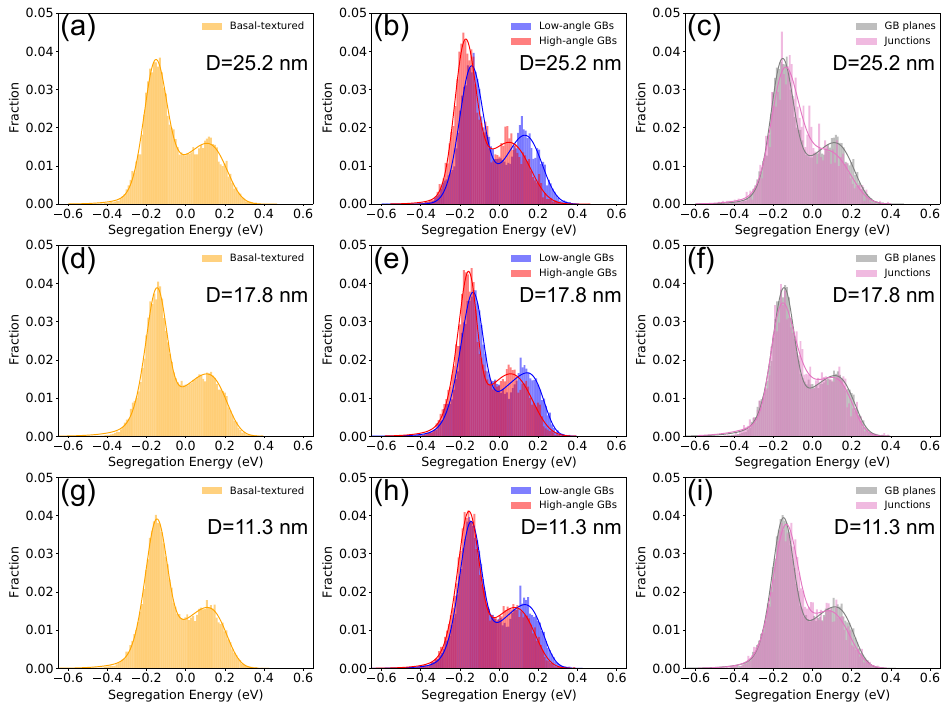}
\caption{Segregation energy spectra of Al solutes at GB in basal-textured nanocrystalline Mg. The spectra are shown for nanocrystalline Mg with different average grain sizes ($D$): (a-c) 25.2 nm, (d-f) 17.8 nm, and (g-i) 11.3 nm. The spectra are further categorized as follows: (a,d,g) all GB sites, (b,e,h) low-angle and high-angle GB sites, and (c,f,i) GB sites with only two neighboring grains and GB junction sites with more than two neighboring grains.The bin size is 0.01 eV.}
\label{figS5}
\end{figure*}

\begin{table*}
\caption[]{\label{tab2} Summary of the parameters used for grain segmentation using Or\textsc{isodata} for basal-textured polycrystals.}
\begin{tabular}{p{0.2\textwidth}p{0.15\textwidth}p{0.15\textwidth}p{0.15\textwidth}p{0.25\textwidth}}
\hline\hline
\addlinespace[0.1cm]
 Grain size $D$ (nm) & Cutoff radius (\AA) & Split threshold & Merge threshold & Minimal grain size (atoms)\\
\addlinespace[0.1cm]
\hline

25.2 & 6.0 & 0.8 & 0.1 & 15000 \\
17.8 & 6.0 & 0.75 & 0.1 & 11000 \\
11.3 & 6.0 & 0.6 & 0.1 & 3000 \\

\addlinespace[0.1cm]
\hline \hline
\end{tabular}
\end{table*}

\begin{table*}
\caption[]{\label{tab1}Summary of fitting parameters, including location ($\xi$, in eV/atom), scale ($\omega$, in eV/atom), shape ($\alpha$, dimensionless), and scaling factor ($\beta$, dimensionless) parameters for bimodal and skew-normal distributions applied to segregation energy $\Delta E_{\text{seg}}$ at 0 K in a randomly-oriented polycrystal and basal-textured ($z$-$\langle 0001 \rangle$) polycrystals with different average grain sizes.}
\begin{tabular}{p{0.25\textwidth}p{0.06\textwidth}p{0.06\textwidth}p{0.06\textwidth}p{0.06\textwidth}p{0.06\textwidth}p{0.06\textwidth}p{0.06\textwidth}p{0.06\textwidth}}
\hline\hline
\addlinespace[0.1cm]
 & $\xi_{1}$ & $\omega_{1}$ & $\alpha_{1}$ & $\beta_{1}$ & $\xi_{2}$ & $\omega_{2}$ & $\alpha_{2}$ & $\beta_{2}$ \\
\addlinespace[0.1cm]
\hline
\addlinespace[0.1cm]
Randomly-oriented & -0.187 & 0.146 & 1.506 & 0.010 & - & - & - & - \\
\addlinespace[0.1cm]
\hline
\addlinespace[0.1cm]
All, $D$=25.2 nm  & -0.126 & 0.065 & -0.715 & 0.005 & 0.209 & 0.230 & -3.631 & 0.005 \\
LAGB, $D$=25.2 nm & -0.104 & 0.074 & -1.015 & 0.005 & 0.210 & 0.153 & -2.224 & 0.005 \\
HAGB, $D$=25.2 nm & -0.173 & 0.053 & -0.007 & 0.005 & 0.152 & 0.201 & -2.502 & 0.005 \\
GB site, $D$=25.2 nm  & -0.124 & 0.066 & -0.819 & 0.005 & 0.210 & 0.226 & -3.616 & 0.005 \\
GB junction, $D$=25.2 nm & -0.184 & 0.083 & 0.967 & 0.005 & 0.187 & 0.236 & -3.048 & 0.005 \\
\addlinespace[0.1cm]
\hline
\addlinespace[0.1cm]
All, $D$=17.8 nm  & -0.110 & 0.071 & -1.365 & 0.004 & 0.208 & 0.244 & -4.08 & 0.006 \\
LAGB, $D$=17.8 nm & -0.091 & 0.083 & -1.767 & 0.005 & 0.229 & 0.228 & -4.832 & 0.005 \\
HAGB, $D$=17.8 nm  & -0.120 & 0.072 & -1.901 & 0.004 & 0.165 & 0.222 & -2.901 & 0.006  \\
GB site, $D$=17.8 nm  & -0.109 & 0.073 & -1.360 & 0.004 & 0.213 & 0.250 & -4.476 & 0.006 \\
GB junction, $D$=17.8 nm & -0.202 & 0.087 & 1.665 & 0.004 & 0.203 & 0.262 & -4.729 & 0.006 \\
\addlinespace[0.1cm]
\hline
\addlinespace[0.1cm]
All, $D$=11.3 nm  & -0.113 & 0.070 & -1.147 & 0.005 & 0.208 & 0.240 & -4.185 & 0.005 \\
LAGB, $D$=11.3 nm & -0.107 & 0.072 & -1.268 & 0.005 & 0.223 & 0.226 & -4.280 & 0.005 \\
HAGB, $D$=11.3 nm  & -0.118 & 0.071 & -1.358 & 0.005 & 0.181 & 0.227 & -3.519 & 0.005 \\
GB site, $D$=11.3 nm  & -0.113 & 0.070 & -1.231 & 0.005 & 0.210 & 0.239 & -4.294 & 0.005 \\
GB junction, $D$=11.3 nm & -0.097 & 0.078 & -1.116 & 0.005 & 0.186 & 0.228 & -3.199 & 0.005 \\
\addlinespace[0.1cm]
\hline\hline
\end{tabular}
\end{table*}

\FloatBarrier
\clearpage

\end{document}